\documentclass[preprint,12pt,authoryear]{elsarticle}
\usepackage{amssymb}
\usepackage{listings}

\journal{...}

\usepackage{color}
\usepackage[normalem]{ulem}
\usepackage[export]{adjustbox}
\usepackage{amsmath}
\usepackage{cases}
\usepackage{subfig}
\usepackage{longtable}
\usepackage{float}
\usepackage[font=footnotesize]{caption}
\usepackage{url,lineno,microtype, eqnarray}
\usepackage[onehalfspacing]{setspace}

\usepackage{bookmark}

\begin{document}
\begin{frontmatter}

\title{FNS: an event-driven spiking neural network simulator based on the LIFL neuron model\\ {\normalsize preprint version}}

\author[label1,label2]{Gianluca Susi\corref{mycorrespondingauthor}} 
\ead{gianluca.susi@ctb.upm.es}

\author[label1]{Pilar Garc\'es} 

\author[label2]{Emanuele Paracone}  

\author[label3]{Alessandro Cristini} 

\author[label3]{Mario Salerno} 

\author[label1,label4]{Fernando Maest\'u} 

\author[label1,label5]{Ernesto Pereda}



\address[label1]{Laboratory of Cognitive and Computational Neuroscience, Center for Biomedical Technology, Technical University of Madrid \& Complutense University of Madrid, Spain.}
\address[label2]{Department of Civil Engineering and Computer Science, University of Rome `Tor Vergata', Italy}
\address[label3]{Department of Electronic Engineering, University of Rome `Tor Vergata', Italy;}
\address[label4]{Department of Experimental Psychology, Cognitive Processes and Logopedy, Complutense University of Madrid, Spain;}
\address[label6]{Department of Industrial Engineering \& IUNE, University of La Laguna, Spain}

\begin{abstract}

Limitations in processing capabilities and memory of today's computers make spiking neuron-based (human) whole-brain simulations inevitably characterized by a compromise between bio-plausibility and computational cost. It translates into brain models composed of a reduced number of neurons and a simplified neuron's mathematical model, leading to the search for new simulation strategies.\\
Taking advantage of the sparse character of brain-like computation, the \emph{event-driven} technique could represent a way to carry out efficient simulation of large-scale \emph{Spiking Neural Networks} (SNN). The recent \emph{Leaky Integrate-and-Fire with Latency} (LIFL) spiking neuron model is event-driven compatible and exhibits some realistic neuronal features, opening new avenues for brain modelling.
In this paper we introduce FNS, the first LIFL-based spiking neural network framework, which combines spiking/synaptic neural modelling with the event-driven approach, allowing us to define heterogeneous neuron modules and multi-scale connectivity with delayed connections and plastic synapses. In order to allow multi-thread implementations a novel parallelization strategy is also introduced.
This paper presents mathematical models, software implementation and simulation routines on which FNS is based. Finally, a brain subnetwork is modeled on the basis of real brain structural data, and the resulting simulated activity is compared with associated brain functional (source-space MEG) data, demonstrating a good matching between the activity of the model and that of the experimetal data.
This work aims to lay the groundwork for future event-driven based personalised brain models.\\

\end{abstract}


\end{frontmatter}


\section{Introduction}
\begin{figure}[!ht]
\centering
\includegraphics[width=1\textwidth]{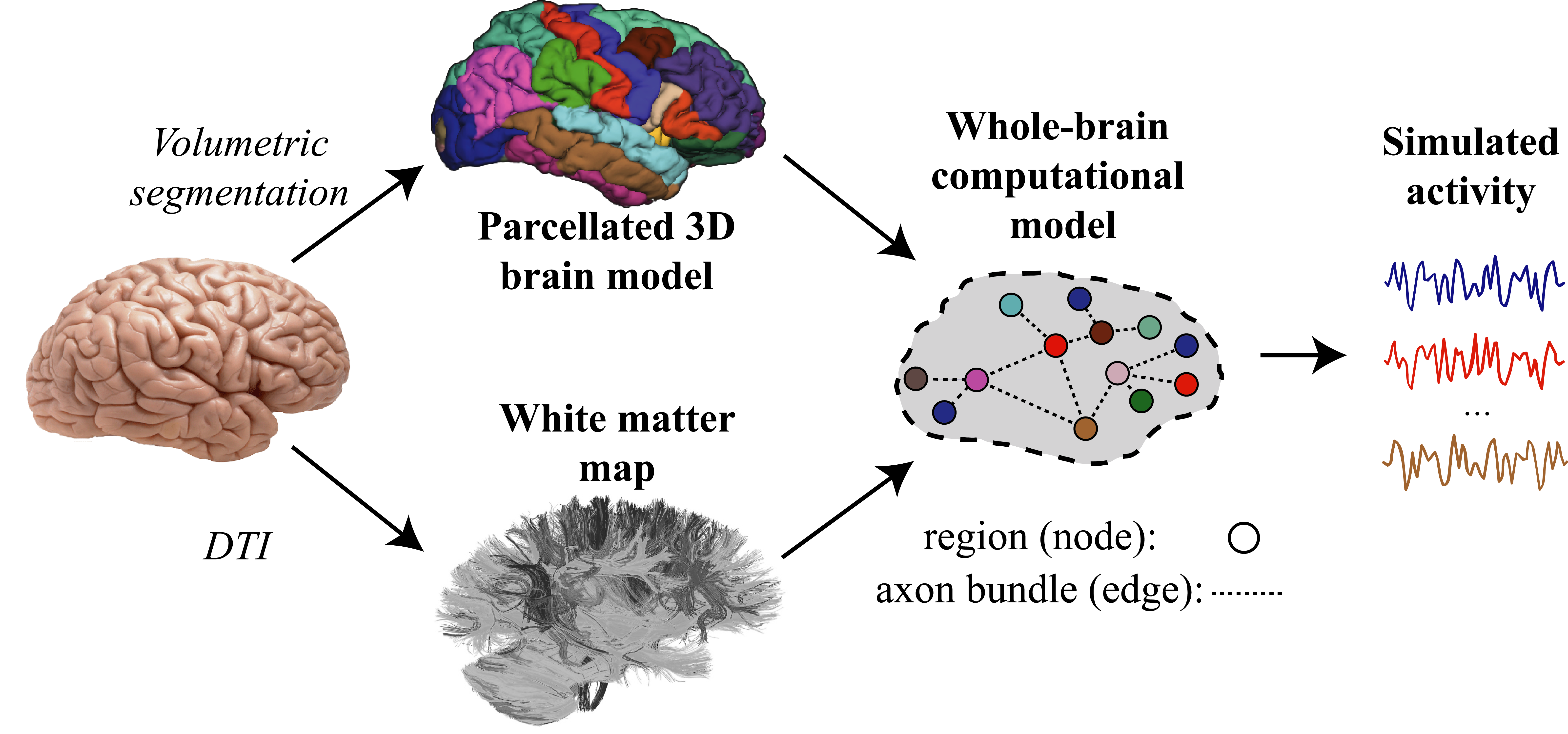}
 \caption{Synthesis of a computational brain model using the graph approach. White matter connections can be extracted by means of DTI. Brains of individual subjects can be coregistered to a parcellation template (\emph{atlas}) in order to assign connections to specific brain areas. By assigning node local dynamics to the obtained \emph{structural connectome}, it is possible to extract simulated activity. The number of nodes of the model depends on the template used, and each node can be represented at different levels of abstraction (e.g., ensemble of spiking neurons).}
\label{FIG:BrainModGrAppr}
\end{figure}

Today's advanced \textit{magnetic resonance imaging} (MRI) -based techniques allow a thorough estimation of the structural \textit{connectome} (i.e., the map of neural connections in the brain \citep{HagmannConn, SpornsConn}), as well as volume and morphology of single brain areas. Through the application of graph theory, such data can be employed to synthesise brain dynamic models, which today are more and more able to appropriately reproduce brain oscillations revealed by functional imaging techniques such as \emph {Multi-Unit Activity} (MUA) and \emph {Local Field Potential} (LFP) \citep{Barardi2014}, \textit{functional} MRI \citep{Cabral2011, Deco2012, Bettinardi2017} and \textit{Magnetoencephalography} (MEG) \citep{Nakagawa2014, Cabral2014, Deco2017}, providing new information on the brain operation.
In such approaches, \emph{nodes} represent surrogates of brain regions (corresponding to \textit{gray matter}), and \emph{edges} represent the long-range connections, along fibre tracts, between them (corresponding to \textit{white matter}), usually estimated using \emph{diffusion tensor imaging} (DTI) (Fig. \ref{FIG:BrainModGrAppr}).

Among the different approaches used to represent brain regions \citep{Deco2008}, \emph{spiking/synaptic models} \citep{Vicente2008, Gollo2010, Nakagawa2014, Maslennikov2014} present a very large number of degrees of freedom, which gives rise to highly complex and realistic behaviours on a broad frequency range of the related oscillations \citep{Barardi2014}. In addition, spiking/synaptic models offer the opportunity to relate to real-brain data transversely \citep [\emph{micro-}, \emph{meso-}, and \emph{macro-scale}, referring to the categorisation of][]{Bohland2009}, as well as to implement \emph{spike-timing dependent plasticity} (STDP), which is indispensable in many kinds of computational neuroscience studies.

On the other hand, spiking/synaptic-based brain simulations are extremely expensive from a computational point of view \citep{Izhikevich2004}. This issue translates to the use of simplified spiking neuron models, and nodes composed of a low number of elements, thereby reducing the realism of the overall brain model.

Spiking neuron models are described by differential equations and usually simulated with clock-driven (synchronous) algorithms, by means of proper integration methods \cite [see][for an extensive review]{Brette2007}. In this way the update is done at every tick of a clock \textbf{X$(t)$ $\rightarrow$ X$(t+dt)$}, and involves all network elements (neurons and possibly synapses). Such two characteristics cause a fast growth of simulation times/used memory when the network size increases, and can lead the simulation to entail missing spikes (although recent work is aimed to overcome the latter limitation, as in \cite{ Hanuschkin2010} and \cite{ Krishnan2017}).

Conversely, in the event-driven (or asynchronous) approach a network element is updated only when it receives or emits a spike. Then, such approach does not envisage a periodic update, neither a check of all network elements, producing simulations devoid of missed spikes, and exploits the sparseness of brain-like activity. Since the latter is irregular in time with low average \citep {Mouraud2009}, this approach has the potential to reduce the computational cost for large-scale network simulations \citep{Ros2006}. Nevertheless, the need of an explicit solution for the neuron state between spikes, and the consideration of incoming and outgoing pulses as discrete events, make the event-driven simulation of classic bio-realistic models very challenging. This has stimulated a big interest in scientific community in developing both realistic and event-driven-compatible spiking neuron models \citep [see][]{Brette2006, Brettepa2007,Tonnelier2007, Salerno2011, RudolphLilith2012}, which led to the development of event-driven based SNN simulators \citep [see][]{Pecevski2014, Cristini2015}, and hybrid \emph{event}/\emph{time-step} based simulation strategies \citep [see][]{Morrison2006, Hanuschkin2010, DHaene2014, NEST2007, BrianDocumentation}. 

In particular, the \emph{Leaky Integrate-and-Fire with Latency} (LIFL) model  is a recent neuron model that can be simulated in event-driven fashion, preserving important computational features at the same time \citep{Susi2018, Salerno2011, Cardarilli2013, Cristini2015, Susi2015, Susi2016, Acciarito2017}. Differently from the \emph{Leaky Integrate-and-Fire} (LIF), LIFL incorporates important neuronal features extracted from the bio-realistic \emph{Hodgkin-Huxley} (HH) model, such as the \emph{spike latency} \citep{FitzHugh1955, Izhikevich2004}. The latter has been proved to be fundamental in many scenarios of neural computation, providing a large range of behaviors. Then, the LIFL may open new avenues for the efficient simulation of large scale brain models.

In this work we present FNS (literally, \emph{Firnet NeuroScience}), a LIFL-based exact event-driven SNN framework oriented to brain simulations, implemented in \emph{Java}. FNS allows us to generate brain network models on the basis of a versatile graph-based multi-scale neuroanatomical connectivity scheme, allowing for heterogeneous neuron modules and connections. In addition to the high-customizability of the network, proper input and output sections make it possible to relate model activity to real data, with the option to enable plasticity, then making the network parameters evolve depending on the network activity. 

In section 2, we describe the neurobiological principles and mathematical models on which FNS is based: neuron model, region model, fibre tracts model, plasticity, input and output signals.
\\
In section 3, we present the possibilities that the framework offers for the synthesis of custom models and the design of specific simulations: generator section, neuroanatomical model section and output section.\\
In section 4, we illustrate the technical aspects of the simulation framework itself: design principles, data structures and parallelization strategy.\\
In section 5, we present an example to show how to conduct a simulation in FNS, and to evaluate the realism and performances of the framework itself. In short, we model a brain subnetwork using structural data of a real subject, and through FNS we simulate brain activity and synthesize electrophysiological-like output signals. Then, we compare such signals with those of the real subject.\\ 
In the Discussion section, we summarize our work and envisage how to improve FNS in future works.

In this manuscript, a single neuron is indicated with $n$; an axonal connection between two whichever neurons with $e$; a neuron module (corresponding to a region or subregion in real case) with $N$, and called \emph{network node}; the complete set of connections between two nodes (corresponding to fibre tracts in real case) with $E$, and called \emph{network edge}.

The software can be freely downloaded at:\\
\url{www.fnsneuralsimulator.org}.\\
In the download link a user guide (including a short description of how to install and run it) and some network models are also provided with the software.

\section {From neurobiology to mathematical models}
Specificity and heterogeneity characterize the human brain at all scales. In this regard, recent works highlight crucial aspects that have to be taken into account in brain models to obtain realistic dynamics:

\begin{itemize} 
\item
\emph{Region bioplausibility}: in spiking/synaptic models, an inappropriate choice of the spiking neuron model or the intra-module connectivity configuration may lead to results having nothing to do with the information processing of real brain \citep{Izhikevich2004}. Of course, also the cardinality of the nodes is important for achieving an appropriate network behaviour.
\item 

\emph{Region diversity}: diversity among and within regions specializes the behaviour of single parts of the network, enhancing the information content and coding performances and shaping properties of collective behavior such as synchronization \cite[see][]{Thivierge2008, Gollo2016}.

\item 
\emph{Inter-region connection bioplausibility}: synchronization between network nodes is strongly sensitive to edge parameters (as weights, delays and connection number) and their distributions \citep{Brunel1999, Brunel2003, Vicente2008, Viriyopase2012, Gollo2014}.

\item 
\emph{Inter-region connection diversity}: selective variations of edge parameters are able to reconfigure the network synchronization profile \citep{Abuhassan2014}, including synchronization between nodes that are not directly connected to the modified edge \cite{Gollo2014}.
\end{itemize}

FNS aims to guarantee the possibility to take into account such aspects in order to avoid the alteration of the network operation. In this section we present mathematical models used in FNS.

\subsection{LIFL Neuron model}
 
Altough the classic LIF model is very fast to simulate, it has been regarded as unrealistically simple, thereby incapable of reproducing the dynamics exhibited by cortical neurons \citep {Izhikevich2003}. FNS is based on the LIFL, that besides being computationally simple it is also able to support a greater number of neuronal features than the LIF.

\subsubsection{A brief introduction to the spike latency neuro-computational feature}
\label{SECT:SLat}
The \emph{spike latency} is the membrane potential-dependent delay time between the overcoming of the ``threshold'' potential and the actual spike generation \citep{Izhikevich2004}. Among all the neuron features, it is of considerable importance because it extends the neuron computation capabilities over the ``threshold'', giving rise to a range of new behaviors.
Spike latency is ubiquitous in the nervous system, including the auditory, visual, and somatosensory systems \citep{Wang2013, Trotta2013}.

From a computational point of view it provides a spike-timing mechanism to encode the strength of the input \citep{Izhikevich2007} conferring many coding/decoding capabilities to the network \citep[e.g.,][]{Gollisch2008, Fontaine2009, Susi2015}, whereas, from a statistical point of view it results in a desynchronizing effect \citep{Salerno2011, Cardarilli2013}, fostering the emergence of higher frequencies \citep{Susi2016} and providing robustness to noise  to the network \citep [][chapter 7]{Izhikevich2007}.
Spike latency has already been introduced in some variants of the LIF, as \emph{QIF} \citep{Vilela2009} and \emph{EIF} \citep{Forcaud2003}. In LIFL spike latency is embedded with a mechanism extracted from the realistic HH model \citep{Salerno2011}, both simple and suitable to the event-driven simulation strategy.

\subsubsection {LIFL operation} 
\label{SECT:LIFLoperation}
In this section, we briefly describe the behaviour of the LIFL neuron model. For the sake of simplicity, we will refer to its basic configuration.

LIFL neuron model is characterized by a real non-negative quantity $S$ (the \emph{inner state}, corresponding to the membrane potential of the biological neuron), which is defined from 0 (corresponding to the resting potential of the biological neuron) to $S_{max}$ (\emph{maximum state}), a value much greater than one, at most $\infty$. 
Simple Dirac delta functions (representing the {action potentials}) are supposed to be exchanged between network's neurons, in form of \emph{pulse} trains.
The model is able to operate in two different modes: \emph{passive mode} when $S<S_{th}$, and \emph{active mode} when $S\geq S_{th}$, where $S_{th}$ is the \emph{firing threshold}, a value slightly greater than 1. 
In passive mode, $S$ is affected by a decay, whereas the active mode is characterized by a spontaneous growth of $S$. Assuming that neuron $n_j$ (i.e., the \emph{post-synaptic neuron}) is receiving a pulse from neuron $n_i$ (i.e., the \emph{pre-synaptic neuron}), its inner state is updated through one of the following equations, depending on whether $n_j$ was in passive or in active mode, respectively:

 \begin{subnumcases}{S_{_j}= }
S_{p\;_{j}}+A_{_{i}}\cdot W_{_{i,j}}-T_l \; , \mbox{ for } 0\leq S_{p\;_{j}}<S_{th} \label{EQ:Pm}
\\
S_{p\;_{j}}+A_{_{i}}\cdot W_{_{i,j}}+T_r \; , \mbox{ for } S_{th}\leq S_{p\;_{j}}< S_{max} \label{EQ:Am}
 \end{subnumcases}
 
$S_{p\;_{j}}$ represents the post-synaptic neuron's \emph{previous state}, i.e., the inner state immediately before the new pulse arrives.  $A_{_{i}}$ represents the \emph{pre-synaptic amplitude}, which is related to the pre-synaptic neuron, and can be positive or negative depending on whether the neuron sends excitatory or inhibitory connections, respectively.
 
$W_{_{i,j}}$ represents the \emph{post-synaptic weight} related to the pre-/post-synaptic neuron couple; if this quantity is equal to 0, the related connection is not present. The product $A_{i}\cdot W_{i,j}$ globally represents the amplitude of the pulse arriving to the post-synaptic neuron $n_j$  (i.e., the \emph{synaptic pulse}) from the pre-synaptic neuron $n_i$.
In this paper, $w$ or $\omega$ will be used instead of W, depending on the connection is intra- o inter- node, respectively.

$T_l$ (the \emph{leakage term}) takes into account the behaviour of $S$ during two consecutive input pulses in passive mode. The user is allowed to select among two kinds of underthreshold decay: \emph{linear} decay \cite[as in][]{Mattia2000} or \emph{exponential} decay \cite[as in][]{Barranca2014}, which behaviour is modulated by the \emph{decay parameter} D, as explained in the \emph{Appendix A}.

$T_r$ (the \emph{rise term}) takes into account the overthreshold growth acting upon $S$ during two consecutive input pulses in active mode. Specifically, once the neuron's inner state crosses the threshold, the neuron is ready to fire. The firing is not instantaneous, but it occurs after a continuous-time delay, representing the spike latency, that we call \textit{time-to-fire} and indicate with $t_f$ in our model. This quantity can be affected by further inputs, making the neuron sensitive to changes in the network spiking activity for a certain time window, until the actual spike generation. $S$ and $t_f$ are related through the following bijective relationship, called the \textit{firing equation}:

\begin{equation}
  t_f ={\frac{a}{(S-1)}}-b \; \label{EQ:Fe}
\end{equation}

where $a,b \ge 0$. Such rectangular hyperbola has been obtained through the simulation of a membrane patch stimulated by brief current pulses (i.e., 0.01 \emph{ms} of duration), solving the Hodgkin-Huxley equations \citep{Hodgkin1952} in \emph{NEURON} environment \citep{Neuron}, as described in \cite{Salerno2011}. Then, if the inner state of a neuron is known, the related $t_{f}$ can be exactly calculated by means of Eq. \ref{EQ:Fe}. As introduced in \ref{SECT:SLat}, this nonlinear trend has been observed in most cortical neurons \citep{Izhikevich2004}; similar behaviors have been also found by other authors, such as \cite{Wang2013} and \cite{Trotta2013}, using DC inputs.
Conversely to previous versions of LIFL \citep{Cristini2015, Susi2018}, positive constants $a$ and $b$ have been introduced in order to make the model able to encompass the latency curves of a greater number of neuron types; in particular, $a$ allows us to distance/approach the hyperbola to its centre, while $b$ allows us to define a $S_{max}$, conferring a bio-physical meaning to the inner state in active mode (note that if $b=0$, then $S_{max}=\infty$; nevertheless, the neuron will continue to show the spike latency feature).

The firing threshold can be equivalently written as:

\begin{equation}
S_{th} = 1+c \; \label{EQ:Sth}
\end{equation}

where $c$ is a positive value called \emph{threshold constant}, that fixes a bound for the maximum $t_{f}$. According to Eq. \ref{EQ:Sth}, when $S = S_{th}$, the $t_{f}$ is maximum, and equal to:

\begin{equation}\label{EQ:TfM}
t_{f,max} = a/c - b
\end{equation}

where $t_{f,max}$ represents the upper bound of the time-to-fire. As mentioned above, the latter consideration is crucial in order to have a finite maximum spike latency as in biological neurons \citep{FitzHugh1955}. From the last equation, we obtain the restriction $c<a/b$.

\begin{figure}
\centering
\includegraphics[width=0.7\textwidth]{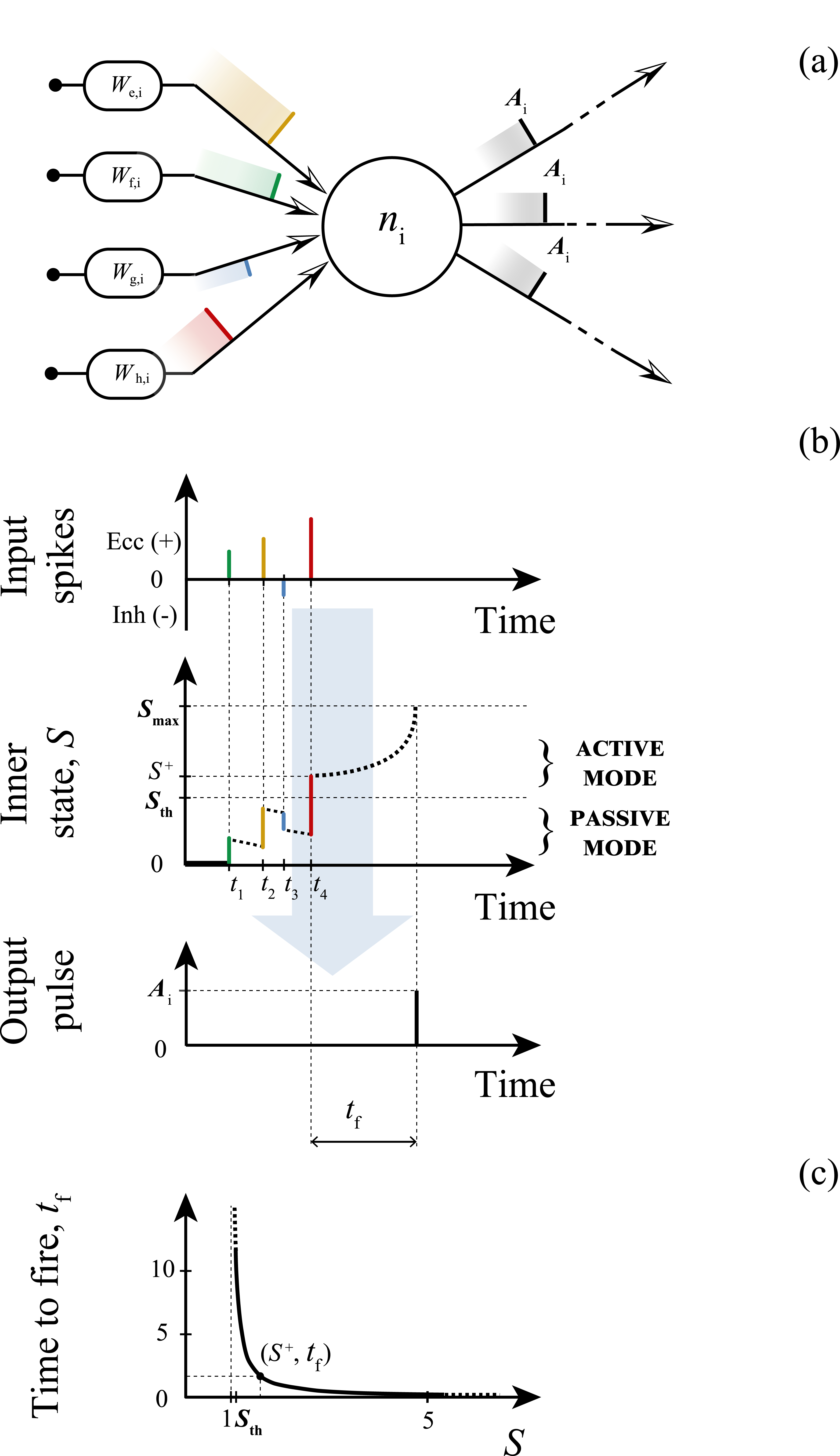}
 \caption{Neural summation and spike generation in a LIFL neuron. (a) Input/output process scheme; (b) temporal diagram of LIFL operation (basic configuration), assuming the neuron starts from its resting potential. 
Each incoming excitatory (inhibitory) input causes an instantaneous increase (decrease) of the inner state. In passive mode the neuron is affected by a decay; when $S$ exceeds the threshold ($S=S^{+}$) the neuron is ready to spike; due to the latency effect, the firing is not instantaneous but it occurs after $t_{f}$. Once emitted, the pulse of amplitude $A_{i}$ (positive, if the neuron $i$ is excitatory as supposed to be in this case, without loss of generality) is routed to all the subsequent connections. In (c) is shown the firing equation, i.e., the  latency curve for the determination of $t_{f}$ from $S^{+}$\citep[see][]{Salerno2011}. The simplest case of firing equation curve has been chosen ($a=1$, $b=0$), and $c$ set to $0.04$}
\label{FIG:LIFLOp}
\end{figure}

\begin{figure}
\centering
\includegraphics[width=1\textwidth]{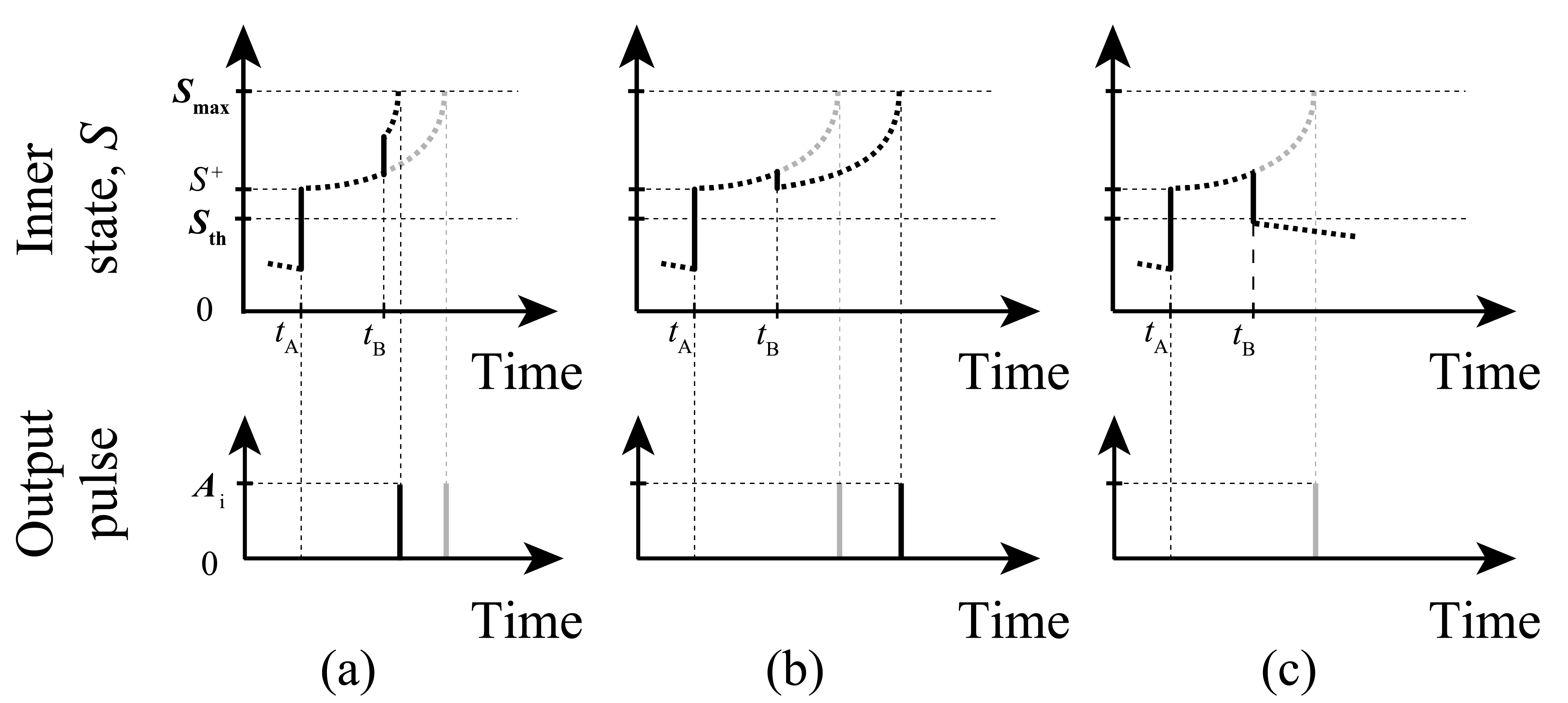}
 \caption{Arrival of further inputs when the neuron is overthreshold. (a) The arrival of a new excitatory synaptic pulse at time $t_{B}$ anticipates the spike generation (post-trigger anticipation). The arrival of a new inhibitory synaptic pulse at time $t_{B}$ is able to (b) delay the spike generation (post-trigger postponement), or (c) to cancel the spike generation (post-trigger inhibition). In order to simplify the comparison, the state evolution in active mode in the simple case of no further inputs is reported in the same figure (grey). Neuron $i$ is supposed to be excitatory as in Fig. \ref{FIG:LIFLOp}.}
\label{FIG:TfModification}
\end{figure}

As described in \emph{Appendix B}, using Eq. \ref{EQ:Fe}, it is possible to obtain $T_r$ (\emph{rise term}), as follows:

\begin{equation}
  T_r =\frac{(S_{p}-1)^{2} \Delta t}{a-(S_{p}-1)\Delta t} \; \label{EQ:Rt}
\end{equation}

in which $S_{p}$ represents the previous state, whereas $\Delta t$ is the temporal distance between two consecutive incoming pre-synaptic spikes. The Eq. \ref{EQ:Rt} allows us to determine the inner state of a neuron at the time that it receives further inputs during the $t_f$ time window. In Fig. \ref{FIG:LIFLOp}, the operation of LIFL is illustrated, while the effect of Eq. \ref{EQ:Rt} is shown in Fig. \ref{FIG:TfModification}.

Assuming that an input spike leads the inner state overthreshold at time $t_{A}$, the arrival of a contribution during the latency time (i.e., at time $t_{B}$) results in a new $t_f$ (i.e., a change of the firing time). Excitatory (inhibitory) inputs increase (decrease) the inner state of a post-synaptic neuron. Therefore, when a neuron is in active mode, excitatory (inhibitory) inputs decrease (increase) the related time-to-fire (\emph{post-trigger anticipation/postponement} respectively). If the inhibitory effect is as strong as to pull the post-synaptic neuron state under the firing threshold, its $t_{f}$ will be suppressed and its state will come back to the passive mode (\emph {post-trigger inhibition})\citep {Salerno2011, Cristini2015}.

For a given neuron $j$ in active mode, the arrival of new input contributions provokes $t_{f}$ updating. Once the $t_{f}$ is reached, the output spike is generated and the inner state is reset.
Note that if incoming spikes are such as to bring $S$ to a value $<0$ ($> S_{max}$), $S$ is automatically put to 0 (a spike is immediately generated). We emphasize the fact that spike latency enables a mechanism to encode neural information, supported from all the most plausible models. Thus, there is lack of information in models that do not exhibit this relevant property.

Hitherto we have discussed a basic configuration of LIFL, which defines an intrinsically \emph{class 1 excitable}, \emph{integrator} neuron, supporting \emph{tonic spiking} and \emph{spike latency}. Nevertheless, thanks to the simplicity of its mathematical model, it can be easily enriched with other neuro-computational features to reproduce different kinds of cortical neurons \cite[see][]{Izhikevich2004} by introducing minimal modifications to the model equations, or by adding extrinsic properties at the programming level. This is the case of \emph{refractory period} for which the neuron becomes insensitive, for a period $t_{arp}$, to further incoming spikes after the spike generation, and \emph{tonic bursting} for which the neuron produces a train of $N_{b}$ spikes interspaced by an interval $IBI$, instead of a single one.

In addition to the spike latency, emerging from the pure computation of the neuron, in the next section another kind of delay will be introduced, independent from the activity, used to characterize the long-range connections between neurons belonging to different groups.

\subsection{Connection between 2 neurons}
\label{SECT:Connection}

In FNS the network nodes are composed of modules of spiking neurons to represent brain regions. Neurons of the same node interact instantaneously, whereas a settable time delay ($\ge 0$) is present between neurons of different nodes to reflect the remoteness between the regions to which they pertain.

\begin{figure}
\centering
\includegraphics[width=0.7\textwidth]{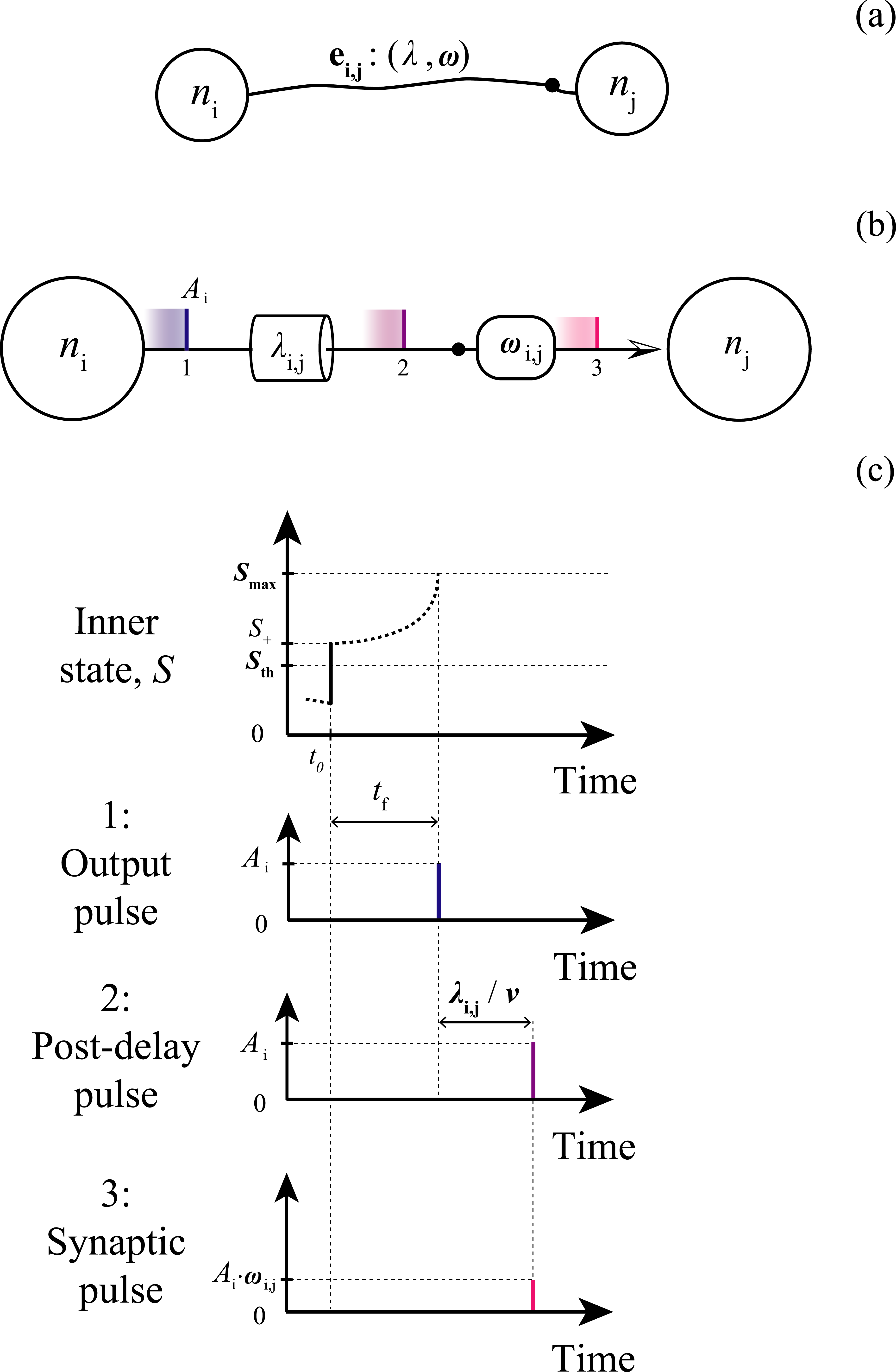}
 \caption{Neuron connection model and pulse transfer. (a) compact representation, (b) logical block representation, (c) temporal diagram: length block produces a translation of the output pulse along time axis. Note that in this example the neuron is supposed to be excitatory, otherwise all the amplitudes would be negative. Output pulses can be considered as a correlate of spiking activity, whereas synaptic pulses can be considered as correlate of synaptic currents. Note that in a) and b) the black dot represents the synaptic junction}
\label{FIG:InterneuronLinkWithDelay}
\end{figure}

\begin{figure}
\centering
\includegraphics[width=0.9\textwidth]{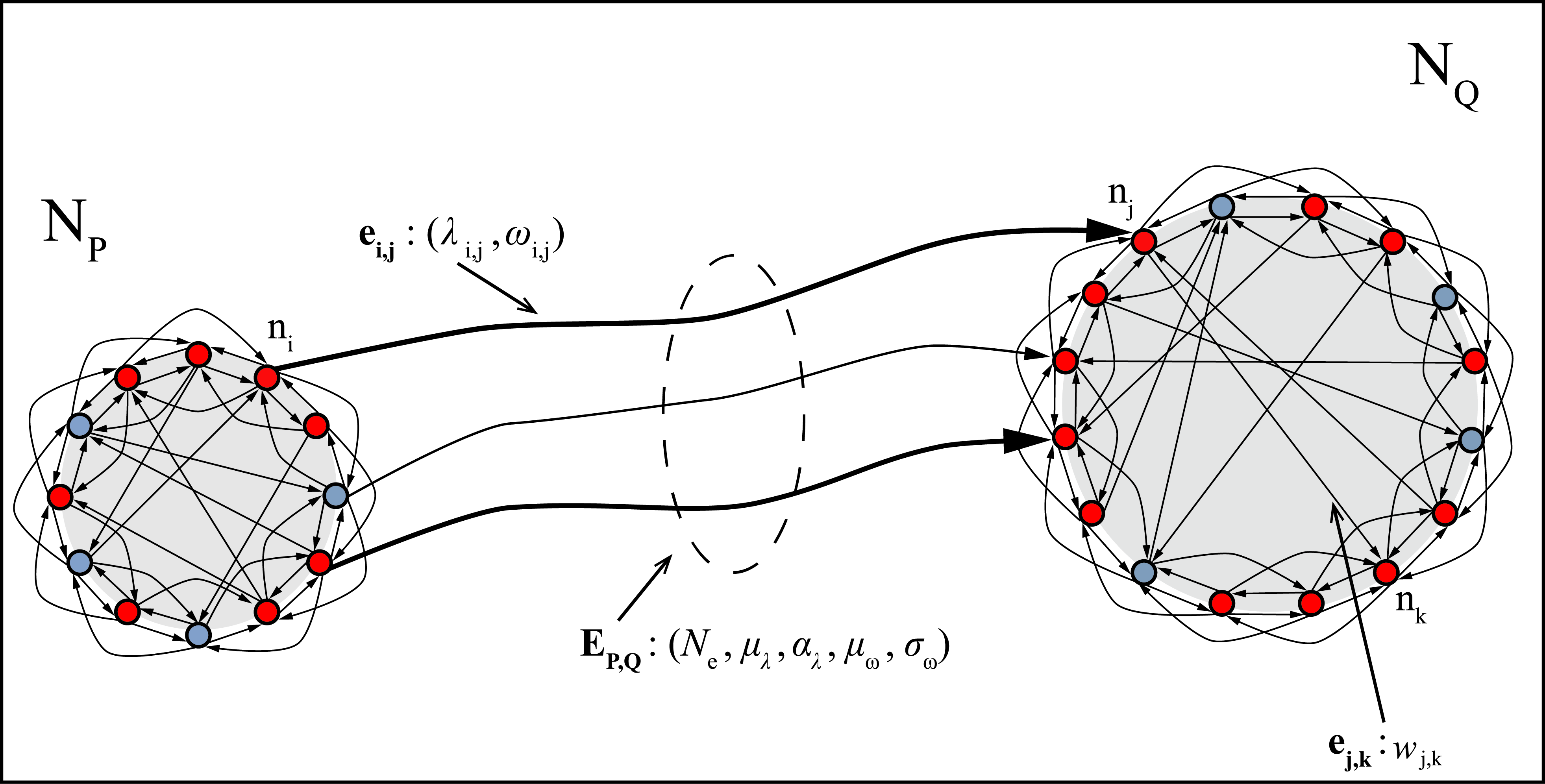}
 \caption{Two nodes connected by an edge.  While an intra-node connection is characterized by its weight, an inter-node connection is defined by weight and length; an edge is defined by number of axons and related distribution of weights and lengths. In order to represent the two modules, for illustrative purpose the following values are used: $n_{1}=12$, $ n_{2}=16$; $R_{1}=2/3$, $R_{2}=3/4$; $k_{1} = k_{2} =4$; $p_{1} =1/6$, $p_{2} =1/8$.}
\label{FIG:NodesBundle}
\end{figure}

A scheme of inter-node neuron connection ($e_{i,j}$) is illustrated in Fig. \ref{FIG:InterneuronLinkWithDelay}, where $\lambda_{i,j}$ represents the \emph{axonal length} block and $\omega_{i,j}$ represents the \emph{post-synaptic weight} block.
Such two link elements (belonging to a directed connection) are able to introduce delay and  amplification/attenuation of the passing pulse, respectively. As in \cite{Nakagawa2014, Cabral2014} a global propagation speed $v$ is set for FNS simulations, so that inter-node connection delays are automatically defined from the axonal lengths, as $\tau_{i,j}=\lambda_{i,j} / v$. Connection delays are important since they allow to take into account the three-dimensionality (i.e., spatial embeddedness) of the real anatomical brain networks.
For the motivations mentioned before, conversely to the inter-node connection (represented as $e_{i,j}$ in Fig. \ref{FIG:NodesBundle}), intra-node connection (represented as $e_{j,k}$ in the same Figure) does not provide the axonal length block (although synaptic weight block continues to be defined).

For biological and mathematical reasons, it is desirable to keep the synaptic weights under a certain value, \emph{$W_{max}$}, a global parameter of the model. 

In the following sections we call \emph{firing event} the pulse emission by a pre-synaptic neuron, and \emph{burning event} the pulse delivery to a post-synaptic neuron.

\subsection{From brain regions to graph nodes}
\label{SECT:BrainRegionNode}

FNS allows us to define regions constituted by one or more \emph{nodes} where each node consists of a neuron module with specific properties. In order to reproduce heterogeneous nodes, a Watts-Strogatz based generative procedure is implemented \citep{Watts1998} as detailed below, allowing the generation of complex networks with structure properties of real neuron populations.

The implemented procedure allows us to model intra- and inter-node diversity: number of neurons and connectivity, percentage of inhibitory neurons, distribution of weights and type of neuron; in addition, it is possible to represent a region with more than one node to model intra-region neuronal pools of different connectivity and neuron types. In the extreme case, a module can be composed of a single neuron, e.g., for reproducing small and deterministic motifs. In the following sections we illustrate the procedure used by FNS for the generation of network nodes and the structure of intra- and inter- node connections.

\subsubsection{Watts-Strogatz-based module generation procedure}
\label{SECT:SWWSproc}

The original Watts-Strogatz procedure is able to generate different types of complex networks (from regular to random), including networks with \emph{small-world} properties (i.e., networks that present large \emph{clustering coefficient} and small \emph{average path length}), that has been demonstrated to reasonably approximate a mid-sized patch of cortex (in the order of $10 \mu m$) with its neighborhood \citep{Riecke2007}. The original Watts-Strogatz procedure is here adapted to generate a module including both inhibitory and excitatory, oriented, connections, analogously to \cite{Maslennikov2014}. 
Given the integer $n$ (i.e., \emph{number of neurons}), $k$ (i.e., \emph{mean degree}), $p$ (i.e., \emph{rewiring probability}), and $R$ (i.e., \emph{excitatory ratio}), with  $0\leq p \leq 1$ and $n\gg k \gg ln(n)\gg 1$, the model generates an oriented graph with $n$ vertices and $nk$ single connections in the following way:

\begin{itemize} 
\item
a regular ring lattice of $n$ spiking neurons is created, of which $R\cdot n$ are able to send excitatory connections and the remaining $(1-R)\cdot n$ are able to send inhibitory connections; 

\item
for each neuron an outgoing connection to the closest $k$ neurons is generated ($k/2$ connections for each side, with $k$ integer and even);  

\item
for each neuron $i$, every link $e_{i,j}$ with $i<j$, is rewired with probability $p$; rewiring is done by exchanging $e_{i,j}$ and $e_{i,m}$ where $m$ is chosen with uniform probability from all possible (excitatory or inhibitory) neurons that avoid self-loops ($m\neq i$) and link duplication. This process is repeated $n$ times, each one considering a different neuron.
\end{itemize} 

Note that the $p$ parameter allows to interpolate between a regular lattice ($p=0$) and a random graph ($p=1$): as $p$ increases, the graph becomes increasingly disordered. For intermediate values of $p$ the network presents small-world properties. The parameters $n$, $k$, $p$ allow the user to customize the network nodes on the basis of the real anatomy. For example, $n$ can be chosen in accord to the volume of the region that is intended to be represented (estimated from a specific subject through volumetry, or extracted from existing \emph{atlases}).

      \subsubsection{Characterization of intra-module connections}

Once connections have been established, weights have to be assigned. Several authors have addressed this problem, setting intra-node weights in different manners. Depending on the specific study, weights have been chosen to have the same, static value \citep{Deco2012}, or characterized by a specific distribution \citep{Abuhassan2014}, or varying in a certain range by means of plasticity \citep{Izhikevich2004b}.
In order to encompass the most of these possibilities, in FNS a set of Gaussian distributed values can be defined by the user for the initialization of the intra-module post-synaptic weights of each module.

\subsection{From fibre tracts to graph edges}

In FSN an \emph{edge} represents a monodirectional set of long-range axons that links a module to another. In the brain, inter-region connections are often characterized by non negligible delays, which are determined by axon length, diameter and myelination.

\subsubsection{Characterization of inter-region connections}

FNS allows the user to set proper number of connections $N_e$ and distributions of weights and lengths can for the network edges. The distribution of edge weights follows a Gaussian function \cite[as in][]{Abuhassan2014}, characterized by the parameters $\mu_{\omega}$ and $\sigma_{\omega}$. Differently, a gamma distribution is implemented for the edge lengths, characterized by mean parameter $\mu_{\lambda}$  and shape parameter $\alpha_{\lambda}$  (see \emph{Appendix C}), since there is probably not a unique prototypical shape for edge delays \cite[as discussed in][]{Vicente2008}. Indeed, this distribution allows the user to explore different shapes, to investigate the impact of different choices on the network activity, to mimic pathological states as the effect of structural inhomogeneity \cite [as discussed in] [] {Ton2014}, or spatially-selective conduction speed decrease due to demyelination \citep {Smith1994}. 

When defining inter-region edges, the user can specify the kind of connections to set among the nodes (excitatory, inhibitory, mixed).

\subsection{STDP} 
\label{SECT:STDP}

Synaptic plasticity consists of an unsupervised spike-based process able to modify weights on the basis of the network activity. The STDP, a well-known type of plasticity mechanism, is believed to underlie learning and information storage in the brain, and refine neuronal circuits during brain development \citep{Sjostrom2010}. Considering a synapse connecting two neurons, such mechanism is based on the precise timings of \emph{pre-synaptic pulse} (i.e., the \emph{synaptic pulse} arriving from the pre-synaptic neuron) and \emph{post-synaptic pulse} (i.e., the \emph{output pulse} generated by the post-synaptic neuron), influencing the magnitude and direction of change of the synaptic weight.
In case of inter-node connection the pre-synaptic pulse is taken after the axonal delay block and not before, in order to not to alter information on causality between pulse arrival and pulse generation.
The original STDP behaviour \cite{Bi1998} can be approximated by two exponential functions \citep{Abbott2000}.

\begin{figure}
\centering
\includegraphics[width=0.9\textwidth]{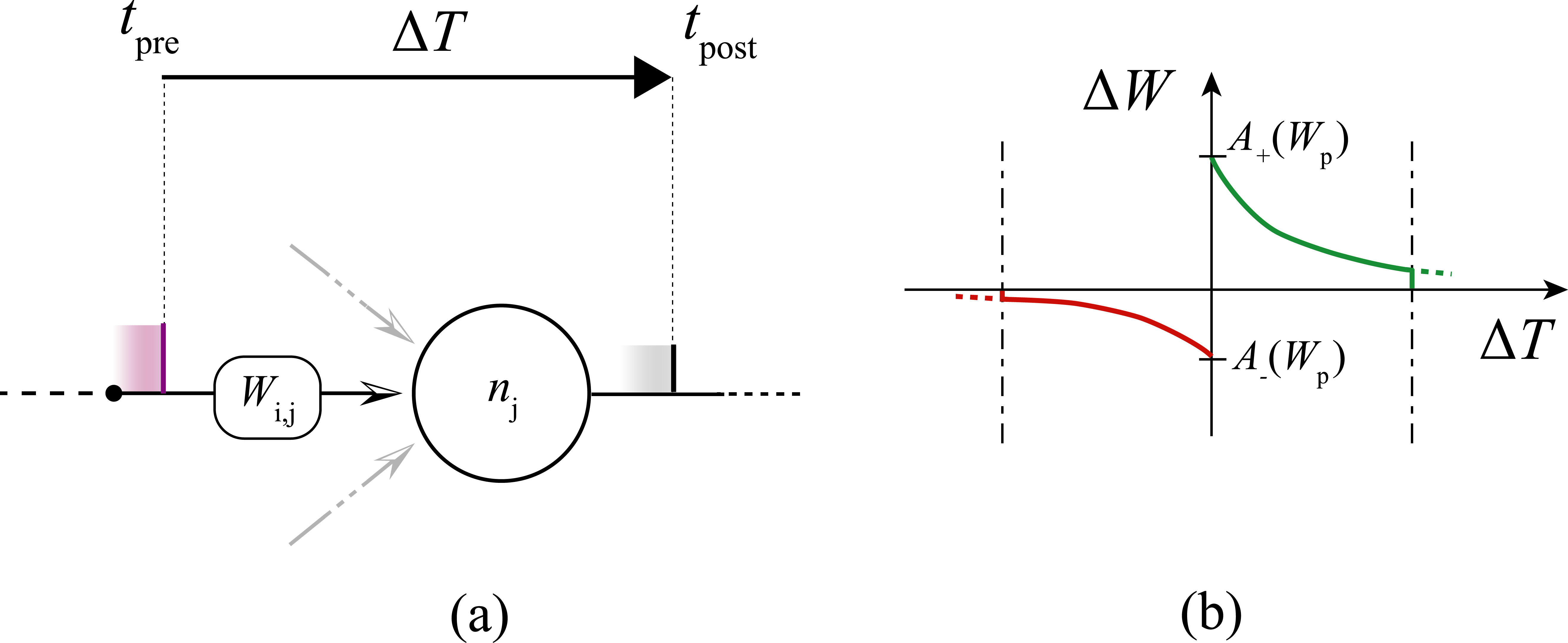}
 \caption{STDP process in FNS: (a) $\Delta T$ calculation in relation to the synapse $w_{ij}$, considering an inter-module connection (without loss of generality); (b) Shapes of the learning windows (LTP in green, LTD in red) considering exponential tail suppression (dash dot).
 }
\label{FIG:STDP}
\end{figure}

 \begin{subnumcases} {\Delta W = }
 \label{EQ:STDP}
 A_+ \mathrm{e}^{-\frac{\Delta T}{\tau_+}} , \mbox{ for } \Delta T > 0\label{EQ:stdpPot}
 \\
 0 , \mbox{ for } \Delta T = 0
 \\
 A_- \mathrm{e}^{\frac{\Delta T}{\tau_-}} , \mbox{ for } \Delta T < 0\label{EQ:stdpDep}
 \end{subnumcases}

where:  
\begin{itemize}
\item
$\Delta T$ is the difference between post-synaptic pulse generation (i.e., $t_{post}$) and pre-synaptic pulse arrival (i.e., $t_{pre}$) instants:
\begin{equation}
\label{EQ:Delta_t}
\Delta T = t_{post} - t_{pre}
\end{equation}
as illustrated in Fig. \ref{FIG:STDP}
\item
$\tau_+$ and $\tau_-$ are positive time constants for \emph{long-term potentiation} (LTP, \ref{EQ:stdpPot}) and \emph{long-term depression} (LTD, \ref{EQ:stdpDep}), respectively;
\item
$A_{+}$ and $A_{-}$ are chosen in order to keep weight values bounded between minimum and maximum values (as discussed in Sect. \ref{SECT:Connection}).
\end{itemize}

Then, weight is increased or decreased depending on the pulse order (\emph{pre-}before \emph{post-}, or \emph{post-} before \emph{pre-}, respectively). To make the weight change dependent also on the current weight value, \emph{soft bounds} \citep{Sjostrom2010} are introduced in FNS, so that $A_+ (W_p)  = (W_{max}-W_{p})\eta_+$ and $A_- (W_p)=W_{p} \eta_-$, where $W_{p}$ is the past value of the synaptic weight, $W_{max}$ the upper bound (see \ref{SECT:Connection}), and $\eta_+$ and $\eta_-$ are positive learning constants, usually in the order of $\sim10^{-5}$. Therefore, the weight update relations implemented in FNS are:

 \begin{subnumcases}{W = }
W_{p} + (W_{max}- W_{p})\eta_+ \mathrm{e}^{-  \frac{\Delta T}{\tau_+}} \;, \mbox{ for }  \Delta T \ge 0 \label{LTP}
\\
W_{p} - W_{p}\eta_- \mathrm{e}^{\frac{\Delta T}{\tau_-}} \;, \mbox{ for }  \Delta T < 0  \label{LTD}
 \end{subnumcases}

It is important to stress that the \emph{soft-bounds} approach allows an increase of both the synaptic capacity and the memory lifetime, with respect to the alternative \emph{hard-bounds} approach \citep{vanRossum2012}.

In addition, to simplify the STDP event list management, exponential tails are suppressed after a certain time value ${TO} \cdot {max}(\tau_+ , \tau_-)$, where \emph{TO} is the \emph{STDP timeout constant}, defined by the user, and usually in the order of 100 ms. In this way, neuron couples whose interval exceeds such time limit are not considered for the STDP process (see Fig. \ref{FIG:STDP}).
 
STDP varies tremendously across synapse types and brain regions \citep{Abbott2000, Caporale2008}.
Accordingly, in FNS it is possible to specify a different set of STDP parameters for each node, or to apply STDP uniquely for certain nodes.

\subsection{Input stimuli}
\label{sec:Input Stimuli}

Several types of stimuli can be of interest in brain simulation studies. Of these, two prototypical types of stimuli are:
\begin {itemize}
\item the noisy fluctuations tipically observed in vivo, which can be modeled by uncorrelated Poisson-distributed spike trains \cite[see][]{Frohlich2008, Abuhassan2014, Nakagawa2014};
\item the DC current used by neurophysiologists to test some neuron features \citep{Izhikevich2004} that can be modeled by constant spike trains \cite[as in][]{Vicente2008}.
\end {itemize}
In addition, in many simulation scenarios the possibility of giving arbitrary spike streams (e.g., sequences that mimic sensory-like processed data) can be of interest, in order to test the response of specific subnetworks.
 
In light of these observations, in FNS it is possible to stimulate brain nodes with three different types of inputs: \emph{Poisson-distributed spike train}, \emph{constant spike train}, and \emph{arbitrary spike stream}. The user is allowed to stimulate all or only a part of the network nodes, choosing for each kind of input a customizable number of fictive excitatory \emph{external neurons}, and the characteristics of the required stimuli.

\subsubsection{Poisson-distributed spike train}
\label{subsec:Poisson Distributed Spikes}

This option provides the injection of spike trains distributed according to an homogeneous Poisson process, in which the underlying \emph{instantaneous firing rate} $r_P$ \citep{Gerstner2014}, is constant over time.

Given a long interval ($t_{A}$, $t_{B}$) we place a single spike in that interval at random. Then, considering the sub-interval ($t_1$, $t_2$) of length $\Delta t = t_2 - t_1$, the probability that the spike occurs during this sub-interval is equal to ${\Delta t}/({t_{B}-t_{A}})$.
Now, placing $k$ spikes in the long interval, the probability that $n$ of them fall in the sub-interval is given by the following binomial formula:

\begin{equation}
P{\{ \mbox{$n$ spikes during $\Delta t$}\}}=\frac{k!}{(k-n)!n!}p^nq^{k-n} \; \label{bin1Formula}
\end{equation}

where $p =\delta t/(t_{B}-t_{A})$ and $q = 1 - p$.
\\

Under proper conditions this expression can be rewritten removing the time dependence as:

\begin{equation}
P{\{ \mbox{1 spike during $\delta t$} \}} \approx r\delta t \; \label{bin2Formula}
\end{equation}

This equation can be used to generate a Poisson spike train by first subdividing time into a set of short intervals, each one of duration $\delta t$. Then generate a sequence of random numbers \textbf{R}, uniformly distributed between 0 and 1. For each interval $i$, a spike is generated if $\textbf{R}(i) \leq r\delta t$. This procedure is appropriate only when $\delta t$ is very small, i.e, only when $r\delta t \ll 1$ \citep{Heeger00poissonmodel}.

In FNS, a user-defined number of fictive \emph{external neurons} $n_{ext P,k}$ is set for each stimulated node $N_k$. By defining a $t_{start P,k}$ and a $t_{end P,k}$ for the external stimuli, each external neuron can send spikes in a discrete number of instants $(t_{start P,k} - t_{end P,k})/ \delta t_P$. The target neurons receive pulses of amplitude $A_{P,k}$.

Pulses are injected from each external neuron to all neurons belonging to a set of nodes defined by the user, by specifying the following set of parameters for each chosen node $N_k$:
$n_{ext P,k}$, $t_{start P,k}$, $t_{end P,k}$, $r_{P,k}$, $\delta t_{P,k}$ and $A_{P,k}$.

\subsubsection{Constant spike train}
\label{subsec:DC input}

This option provides the injection of constant spike trains in order to emulate DC current stimulation. Note that since we simulate the network by means of an event-driven approach, the \emph{DC} input is not continuous as in the real counterpart, but it is constantly sampled with an adequately small time step (i.e., smaller than the spike duration) called \emph{interspike interval} and indicated with $int_{\,c}$. 

In FNS, a user-defined number of fictive \emph{external neurons} $n_{ext \,c,k}$ is set for each stimulated node $N_k$. Each external neuron can send spikes 
from time $t_{start \,c,k}$ to $t_{end \,c,k}$, with amplitude $A_{\,c,k}$.
Such kind of input is injected from each external neuron to all neurons belonging to a set of nodes defined by the user, by specifying the following set of parameters for each chosen node $N_k$: $n_{ext \,c, k}$,$t_{start \,c,k}$, $t_{end \,c,k}$, $int_{\,c,k}$ and $A_{\,c,k}$.

Note that the situation $int_{\,c,k} < t_{arp,k}$ should be avoided because pulses would arrive during the refractory time.

\subsubsection{Arbitrary spike stream}

Arbitrary spike streams can be injected to neurons belonging to a set of nodes defined by the user by specifying the following set of parameters for each chosen node $N_k$: the spike \emph{amplitude} $A_{ss,k}$, and a couple ($n_{ss,k}$, $t_{ss,k}$)  for each event to introduce (i.e., \emph{external source number} and related \emph{spike timing}, respectively). External sources are permanently associated to the neurons of the indicated node, using a random procedure.

\subsection{Output signals}

Depending on the type of contributions we are considering at the network level, i.e., output pulses (corresponding to \emph{action potentials}) or synaptic pulses (corresponding to \emph{post-synaptic currents}), the same network activity gives rise to different signals, due to the presence of connection delays and weights.

In particular, action potentials coincide with the activity emerging from \emph{firing} events (see Sect. \ref{SECT:LIFLoperation}), because they take place before the axon, thus they are spatially localized at the emitter node; whereas post-synaptic currents coincide with the post-synaptic activity (see Sect. \ref{SECT:LIFLoperation}), because they take place downstream the axon, thus they are spatially localized to the receiver node, and are affected by the shifting effect introduced by (heterogeneous) fibre tract's delays and post-synaptic weights.

Action potentials are of interest for some studies \citep [see][]{Vicente2008}, whereas post-synaptic currents can be useful for some others (see \cite{Mazzoni2008, Nakagawa2014} for LFP and MEG signal reconstruction).

In order to give the user the possibility to recostruct such different types of signals, output section of FNS allows to store both pulse emission and arrival times ($t_F$ and $t_B$), transmitter and receiver neurons ($n_F$ and $n_B$) and related nodes ($N_F$ and $N_B$), as well as amplitude weights ($W_{ev}$) involved in each event occurring during the simulation interval, for some nodes indicated by the user before the simulation starts.

\section{Simulation framework structure}

\begin{figure}[h]
\centering
\includegraphics[width=0.95\textwidth]{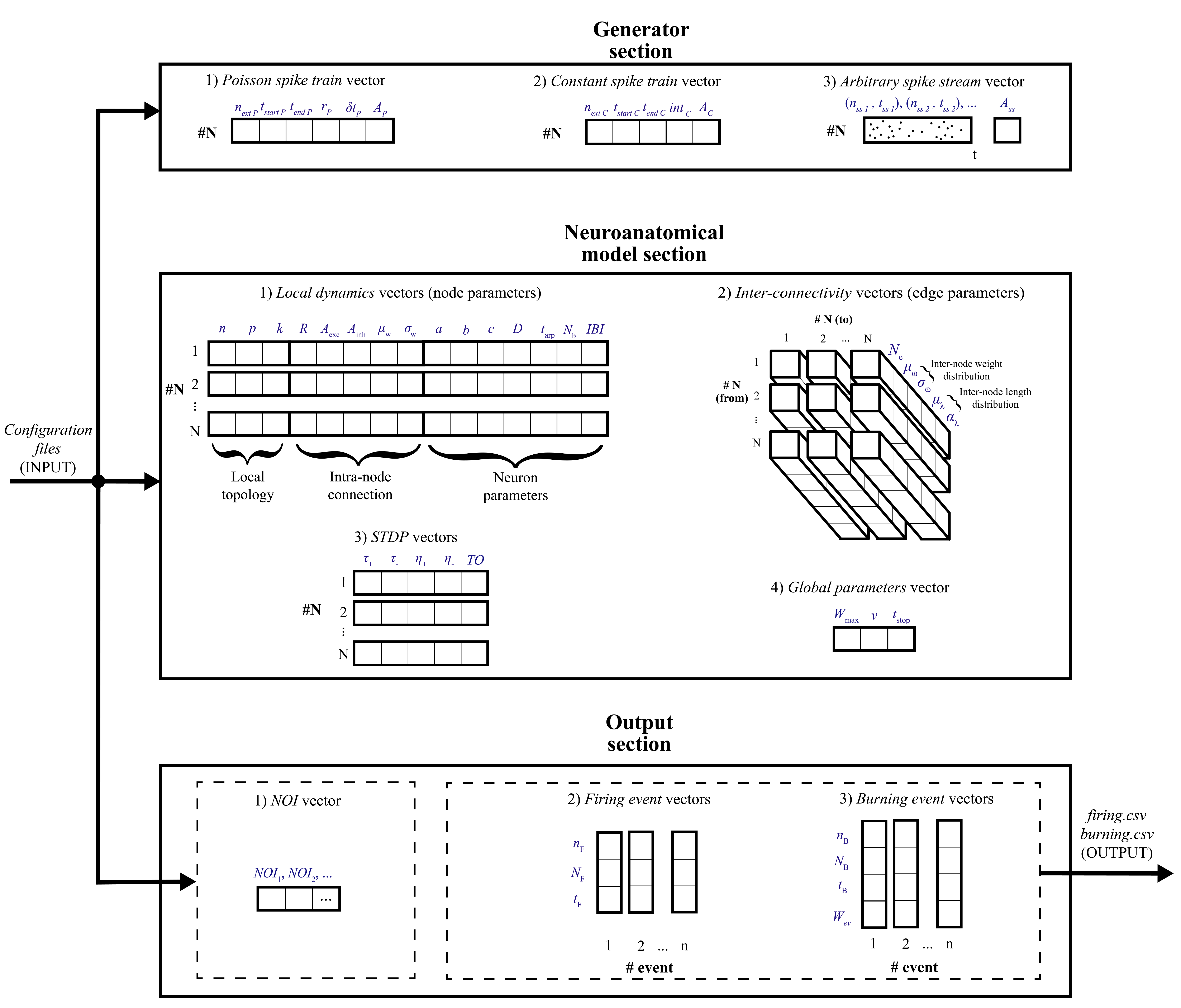}
 \caption{FNS framework overall structure. The reader can find the means of the abbreviations in Table \ref{TAB:ParDef}}
\label{FIG:SIM}
\end{figure}

On the basis of the modelling introduced in the previous section, here we describe the framework structure and the tools it offers to the user for implementing a custom network, stimulating it, and obtaining the outputs of interest.\\
The framework is articulated in three main sections : \emph{Generator section}, \emph{Neuroanatomical model section} and \emph{Output section} (see Fig. \ref{FIG:SIM}). In order to design a simulation, the user interacts with such sections by means of proper configuration files, that are internally translated into configuration \emph{vectors}, which are defined in table \ref{TAB:ParDef}.

{\scriptsize
\begin{center}

\begin{longtable}{|*4{c|}}

\newpage
\hline
\textbf{Section}&\textbf{Vector}&\textbf{Components}&\textbf{Name}\\ 
\hline \hline
Generator&$\mathbf{PV}$&$n_{ext P}$&number of \emph{Poisson spike train} external neurons\\
\cline{3-4}
section&&$t_{start P}$&Poisson input onset\\
\cline{3-4}
&&$t_{end P}$&Poisson input offset\\
\cline{3-4}
&&$r_P$&firing rate\\
\cline{3-4}
&&$\delta t_P$&delta\\
\cline{3-4}
&&$A_P$&Poisson input amplitude\\
\cline{2-4}

&$\mathbf{CV}$&$n_{ext \, c}$&number of \emph{constant spike train} external neurons\\
\cline{3-4}
&&$t_{start{\,c}}$&constant input onset\\
\cline{3-4}
&&$t_{end\,c}$&constant input offset\\
\cline{3-4}
&&$int_{\,c}$&interspike interval\\
\cline{3-4}
&&$A_{\,c}$&constant input amplitude\\
\cline{2-4}

&$\mathbf{SV}$&$t_{ss 1}, t_{ss 2}, ...$&  \emph{input stream} spike timings\\
\cline{3-4}
&&$n_{ss 1}, n_{ss 2}, ...$&related neuron numbers\\
\cline{3-4}
&&$A_{ss}$&stream input amplitude\\
\cline{1-4}

Neuroanatomical&$\mathbf{LDV}$&$n$&number of neurons\\ 
\cline{3-4}
module&&$p$&rewiring probability\\
\cline{3-4}
section&&$k$&mean degree\\
\cline{3-4}
&&$R$&excitatory ratio\\
\cline{3-4}
&&$A_{exc}$&excitatory pre-synaptic amplitude\\ 
\cline{3-4}
&&$A_{inh}$&inhibitory pre-synaptic amplitude\\ 
\cline{3-4}
&&$\mu_{_w}$&intra-node post-synaptic weight distr.mean (Gaussian)\\
\cline{3-4}
&&$\sigma_{_w}$&intra-node post-synaptic weight distr.st.dev. (Gaussian)\\
\cline{3-4}
&&$a$&latency curve center distance\\
\cline{3-4}
&&$b$&latency curve x-axis intersection\\
\cline{3-4}
&&$c$&threshold constant\\
\cline{3-4}
&&$D$&decay parameter\\
\cline{3-4}
&&$t_{arp}$&absolute refractory period\\
\cline{3-4}
&&$N_b$&burst number\\
\cline{3-4}
&&$IBI$&inter-burst interval\\
\cline{2-4}
&$\mathbf{ICV}$&$N_{e}$&number of connections (edge cardinality)\\ 
\cline{3-4}
&&$\mu_{\omega}$&inter-node post-synaptic weight distr.mean (Gaussian)\\
\cline{3-4}
&&$\sigma_{\omega}$&inter-node post-synaptic weight distr.st.dev. (Gaussian)\\
\cline{3-4}
&&$\mu_{\lambda}$&inter-node length distr.mean (gamma)\\
\cline{3-4}
&&$\alpha_{\lambda}$&inter-node length distr.shape (gamma)\\
\cline{2-4}
&$\mathbf{STDPV}$&$\tau_{+}$&LTP time constant\\ 
\cline{3-4}
&&$\tau_{-}$&LTD time constant\\ 
\cline{3-4}
&&$\eta_{+}$&LTP learning constant\\
\cline{3-4}
&&$\eta_{-}$&LTD learning constant\\
\cline{3-4}
&&$TO$&STDP timeout constant\\
\cline{2-4}
&$\mathbf{GPV}$&$W_{max}$&maximum weight\\
\cline{3-4}
&&$v$&global conduction speed\\
\cline{3-4}
&&$t_{stop}$&simulation stop time\\
\cline{1-4}

Output&$\mathbf{NV}$&$NOI_1, NOI_2, ...$&list of \emph{NOI}s\\ 
\cline{2-4}
section&$\mathbf{FV}$&$n_{F}$&pre-synaptic neuron number (if firing event)\\ 
\cline{3-4}
&&$N_{F}$&pre-synaptic node number (if firing event)\\
\cline{3-4}
&&$t_F$&firing event time (if firing event)\\
\cline{2-4}
&$\mathbf{BV}$&$n_{B}$&post-synaptic neuron number (if burning event)\\ 
\cline{3-4}
&&$N_{B}$&post-synaptic node number (if burning event)\\
\cline{3-4}
&&$t_B$&pulse arrival time (if burning event)\\
\cline{3-4}
&&$W_{B}$&synaptic weight (if burning event)\\

\hline \hline
 \caption{Definition of the system parameters.}
 \label{TAB:ParDef}
\end{longtable}
\end{center}
}

\subsection{Generator section}

This section allows the user to inject the desired input to some selected nodes. \emph{Poisson spike train vectors} (i.e., $\mathbf{PV}$), \emph{constant spike train vectors} (i.e., $\mathbf{CV}$) and \emph{arbitrary spike stream vectors} (i.e., $\mathbf{SV}$) can be combined to send more than a kind of input to the same node simultaneously.

\subsection{Neuroanatomical model section}
\label{SECT:NaMS}
This section allows the user to define the network model: local dynamics, structural parameters, plasticity constants and global parameters. 
Each node is fully characterized by a \emph{local dynamics vector} (i.e., $\mathbf{LDV}$), consisting of \emph{local topology parameters}, \emph{intra-node connection parameters} and \emph{neuron parameters}. From the definition of node's weight distribution, the simulator computes all the single intra-node synaptic weights and stores them in proper data structures (see Sect. \ref{SECT:dataStr}).
\\Each edge is fully characterized by a \emph{inter-connectivity vector} (i.e., $\mathbf{ICV}$), consisting of \emph{edge cardinality} and \emph{inter-node weight distribution} and \emph{length distribution parameters}. From the definition of such parameters the simulator computes all the single inter-node lengths and weights and stores them in proper data structures (see Sect. \ref{SECT:dataStr}).

Note that values arising from the distribution of inter-node lengths have to be positive to be stored in the internal matrices; in case they assume negative values FNS allows the user to consider the absolute value of such quantities, or to terminate the program execution. In addition, weight values are kept below the value $W_{max}$.
\\The \emph{STDP vector} (i.e., $\mathbf{STDPV}$) contains all STDP parameters discussed in Sect. \ref{SECT:STDP} and defines the STDP to act on a specific node.
\\The global parameters of the system are defined by the \emph{global parameters vector} (i.e., $\mathbf{GPV}$). Among these, $t_{stop}$ specifies the neural activity time we want to simulate in biological time units ($ms$).

\subsection{Output section}
\label{SECT:OtpS}
\begin{figure}
\centering
\includegraphics[width=0.98\textwidth]{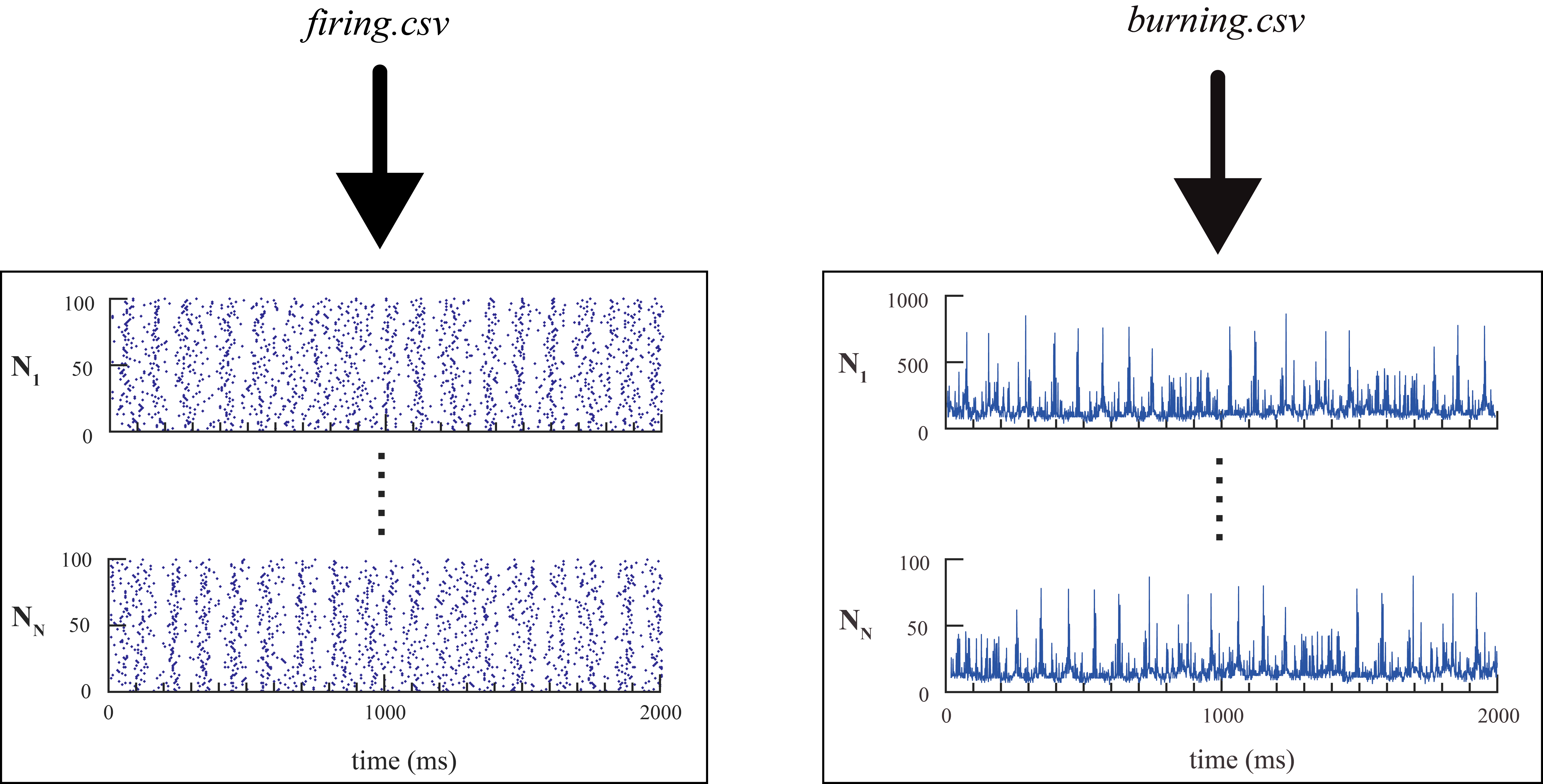}
 \caption{Two seconds of simulated electrophysiological signal extracted from \emph{firing.CSV} and \emph{burning.CSV} files (spike rasterplot and postsynaptic activity, respectively) of a simulation of $N_N$ nodes, composed of 100 neurons each. The figures have been obtained throught the Matlab{\textregistered} script available on the FNS website.} 
\label{FIG:Output}
\end{figure}

This section allows the user to choose regions and type of contributions for which to extract information. Before the simulation starts, the user can specify the list of nodes for which to store all simulation data (\emph{nodes of interests}, or NOIs) through the \emph{NOI vector} (i.e., $\mathbf{NV}$). At the end of the simulation, data of all firing and burning events in which such NOIs are implicated are available to the user, in form of a vector for each event. Depending whether it is a firing or burning event, in the output we obtain different vectors: \emph{firing event vector} (i.e., $\mathbf{FV}$) or \emph{burning event vector} (i.e., $\mathbf{BV}$), respectively. Such vectors are collected and made available to the user through the two files \emph{firing.CSV} and \emph{burning.CSV}. The former reports exhaustive information on firing events and burning events, respectively, to extract simulated electrophysiological signal (firing activity in the first case and postsynaptic activity in the second case), see Fig. \ref{FIG:Output}.

\section{Implementation aspects}

When a simulation is launched in FNS, two phases are performed in sequence: the first phase consists in the \emph{initialization} of the data structures needed by the simulation; the second phase consists in the actual \emph{simulation} of events.

The first phase is carried out through the following steps:
\begin{itemize}
\item reading of the \emph{Generator section} and \emph{Neuroanatomical model section} configuration vectors, and the list of NOIs for which the event vectors \emph{FV} and \emph{BV} (of \emph{Output section}) have to be stored;
\item creation of the node-specific data structures and neuron clusters to be run concurrently;
\item creation of the global data structures.
\end{itemize}

After those steps have been accomplished, the second phase begins. The parallelization strategy implemented allows the program to proceed through the sequential simulation of single slices of simulated time with a specific and constant duration, for multiple nodes at the same time. Each cycle terminates with the synchronization between nodes whose events affects each other.

\subsection {Data structures}
\label{SECT:dataStr}

During the initialization phase the program generates the sets of values of weights and delays of the network. Possible values arising from the distributions that are negative or $> W_{max}$ are notified and rectified (see Sect. \ref{SECT:NaMS}). Below we briefly describe the main data structures used by the software, highlighting which of these act at the node level:

\begin {description}

\item \textbf{inter-connection} \emph{dictionary}: it is a map containing weight and length of each inter-node connection. The connection is identified by the pre-synaptic neuron (pertaining to node $A$) and the post-synaptic neuron (pertaining to node $B$, with $B \neq A$);

\item \textbf{intra-connection} \emph{dictionary} (node-specific): this is the intra-node equivalent of the \emph{inter-connection} dictionary, where each entry represents the weight of an intra-node connection.  The connection is identified by the pre-synaptic neuron and the post-synaptic neuron (pertaining to the same node);

\item \textbf{state} \emph{dictionary} (node-specific): it contains the inner states of the neurons pertaining to a specific node, and it is constantly updated through the node simulation;

\item \textbf{active neuron} \emph{list} (node-specific): list of neurons in active mode pertaining to a specific node, sorted on their firing time; this list is constantly updated through the node simulation;

\item \textbf{outgoing spike} \emph{list} (node-specific): lists of output pulses, including post-synaptic neuron, node and instant, generated from a specific node within a specific time slice;

\item \textbf{STDP timing} \emph{list}: it temporarily stores event timings in order to compute the $\Delta W$. Such timings are automatically discarded after the \emph{TO} value defined by the user.

\end {description}

\subsection{Event-driven procedure}

An asynchronous or event-driven algorithm allows the simulation to "jump" from an event to the next one. If the SNN was characterized by identical transmission delays and absence of spike latency, the data structure would be just a \emph{First In First Out} queue, which has fast implementations \citep{Cormen2001}. Although the neuronal and connectivity features of FNS require more complex data structures (as shown in Sect. \ref{SECT:dataStr}), the simulation procedure is quite simple. \\
At any instant, each network neuron is characterized by its inner state, and active neurons are also characterized by their proper $t_f$. When a firing event occurs, it propagates toward the target neurons taking into account the connection delays (if present). Such events modify the inner state of post-synaptic neurons (and their $t_f$, for the active neurons), on the basis of the amplitude and sign of the pulse, and the time elapsed from the last state update. Four different cases of state update can happen to the target neuron:

\begin{description}
\item [$\cdot$] \emph{passive-to-passive}. This does not have any effect on the event list
\item [$\cdot$] \emph{passive-to-active}. This elicits the insertion of an event (and related firing time) orderly on the event list 
\item [$\cdot$] \emph{active-to-active} (i.e., post-trigger anticipation/postponement). This elicits the update (and reordering) of its firing time on the event list
\item [$\cdot$] \emph{active-to-passive} (i.e., post-trigger inhibition). This elicits the elimination of an event (and related firing time) from the event list
\end{description}

In addition to the four cases listed, two ``forbidden'' cases can occur during the simulation: from \emph{passive mode} to $S<0$ and from \emph{active mode} to $S\ge S_{max}$; for such cases, specific actions are included in the procedure (state value correction and output spike production, respectively).

At the same time, weights for which plasticity is active are updated accordingly, taking in account the \emph{STDP timing list}.

\subsection{Parallelization}

In a parallel computing scenario the problem is splitted in many sub-problems such that their solutions can be computed independently and then collected to provide the global solution. 
In simulations of brain networks it is not trivial to determine which events can be executed in parallel because of the intricate cause-effect relations between the network elements. This led to the development of specific strategies for parallelising event-driven SNNs avoiding causality errors \citep [e.g.][]{DHaene2006, Mouraud2009, Lobb2005, Grassmann1998, Djurfeldt2005, DelormeSPIKENET2003}.\\
The event-driven parallelization method on which FNS is based on can be defined as \emph{adaptive} \citep{DHaene2006}, since the algorithm chooses an appropriate network-specific interval of simulated time to be used for the synchronization of the parallel tasks, avoiding as much as possible the underuse of available hardware resources. 
Given a generic network, the \emph{opaque period} (OP) is the minimum simulated time needed by a signal to travel from a network element to an adjacent one \citep{Lubachevsky1989}. Then, within any simulated time window smaller than the OP of the network, each event cannot be caused by (or cannot affect to) any other event happened during the same time window. 

If $ev_i$ and $ev_j$ are two distinct events such as 

\begin{equation}
\left|\tau(ev_j)-\tau(ev_i)\right| < OP, \quad i \neq j
\end{equation}
\\
then they can be computed in parallel without loss of cause-effect relationship.
This allows us to parallelize the computation within the \emph{time slice} $T_s< OP$.
On the other hand, each unit must wait until each of the others has ended to simulate the events of the previous $T_s$ to process new information; then, a \emph{sync step} is needed to "deliver" the events just calculated to the unit which should use them to produce new events during the next \emph{OPs}.

In the case of neural computation, an event could affect another one in a very short time, leading to a short $OP$, then counteracting the benefit of the parallelization. In order to efficiently perform parallel computation, in FNS the following strategy is adopted:
\begin{itemize}
\item each node is assigned to a specific \emph{thread} (i.e., a process that deals with a local problem);
\item the $T_s$ duration is sized as the minimum among all the network inter-node connection delays (that we define \emph{Bounded Opaque Period} (BOP)), since this is the shortest interval needed by a neuron of a node to affect the state of a neuron of another node.
\item inter-node spikes in queue are delivered to the corresponding threads through the \emph{synch step}, at intervals equals to the BOP. Once a thread gets a spike event from a node, it puts this event orderly to the internal node-specific \emph{active neuron list} and updates the internal state of the post-synaptic neuron at the proper time.
\end{itemize}

If the network presents two or more inter-connected nodes with zero-delay, FNS considers such nodes as a single \emph{group}, and the mechanism continues to be valid. The fact of considering the more ample concept of group instead that of node enables the possibility of representing heterogeneous regions without losing the parallelization feature (see Sect. \ref{SECT:BrainRegionNode}). 
 
Threads are executed in parallel by the \emph{thread workers} (i.e., multi-threaded or hyperthreaded CPUs), each of which can execute at most a thread at a time. In order to minimize the processing times, each worker can serve queued threads in turns for a short time, and different workers can swap threads each other, to achieve a dynamic balancing of the computational load.

Once $\mathbf{ICV}s$ (and then, the groups) have been defined, we can simulate the neural activity within each group. Each \emph{thread} can assume one of the three states \emph{running}, \emph{waiting} or \emph{runnable}.
When the whole simulation starts, the steps until the completion of the execution are the following:\\

\begin{enumerate}
\item all threads are set as \emph{runnable}; \\
\item each of them simulates the generation of events within the (simulated)  $BOP_1$ (i.e., the time window from time $t=0$ to time $t=\bar{B}$).
If empty threads occur, for them the simulation jumps directly to point 3; \\
\item once a thread has generated all the events within the current $BOP_1$ window, it sets its status to \emph{waiting};\\
\item once all threads have completed to simulate the events of the $BOP_1$ window, all threads synchronize each other through the synch step;\\
\item all threads get again to the state \emph{runnable} and simulate the events generated within the $BOP_2$ (i.e., the time window from time $t=\bar{B}$ to time $t=2\bar{B}$), and so on. \\

\end{enumerate}

This algorithm cycles until the stop condition set on the overall simulated time $t_{stop}$.\\
Obviously, a firing event generated within ${BOP_n}$ not necessarily will be delivered as burning event during ${BOP_{n+1}}$: it could be delivered in one of the following BOPs, depending on the connection delay involved.  \\

\begin{figure}[h]

\centering
\includegraphics[width=1\textwidth]{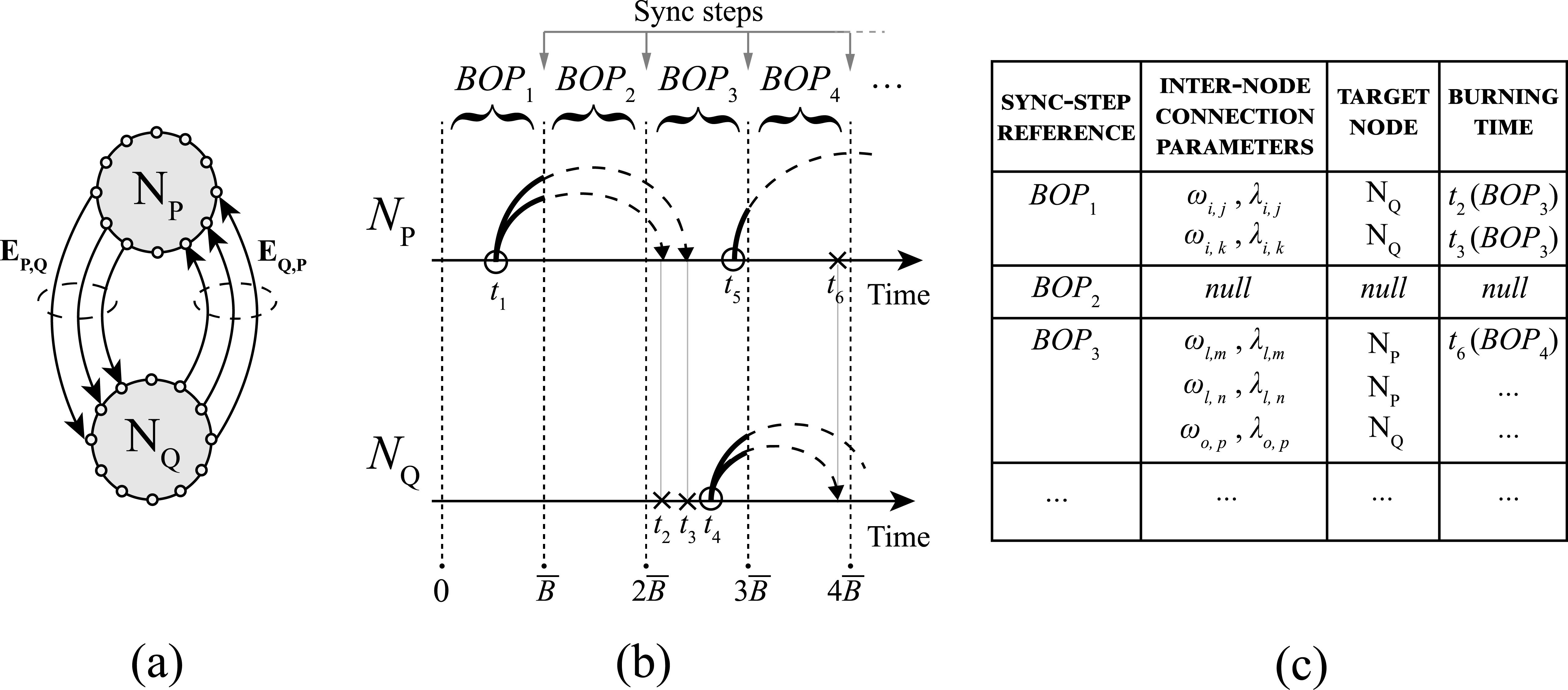}
\caption{BOP-based parallelization mechanism. (a) Example network: two interconnected nodes are generating spiking activity (under proper external stimulation, not shown in the figure). (b) Temporal diagram of the parallelization process. Inter-node pulses are represented by dashed arrows, firing events by circles, burning events by crosses. An inter-node firing event generates burning events in the target node. The delivery of the events to the burning node will happen at the end of the firing event's BOP. Given a firing event, related burnings will happen in one (or more) following BOPs. Given a burning event, the related firing event belongs to a BOP that precedes the current one (not necessarily the previous one). The axis represents simulated time. (c) Event schedule grid.}
\label{FIG:BOP}
\end{figure}

The concept of parallelization through BOP is summarized in Fig. \ref{FIG:BOP}, where for simplicity we consider the simple \emph{resonance pair} motif \citep{Gollo2014, Maslennikov2014} (i.e., two nodes bidirectionally connected with delay).

In \emph{Appendix D} we report a description in pseudo-code of the procedure implemented in FNS, supporting the synchronization mechanism between nodes.

\section{Reproduction of spontaneous MEG functional connectivity}
\label{SECT:ReprFC}

In order to reproduce subject specific brain's spontaneous electrophysiological activity using FNS, both structural and functional \emph{connectomes} (i.e., the maps of structural and functional connections within brain areas \citep{Fornito2016}) have been extracted from a healthy participant using DTI and source-space MEG, respectively. Connectomes have been estimated using 14 regions (7 regions per hemisphere, see table \ref{TAB:NodDef}) composing the Default Mode Network \citep{Raichle2001, DePasquale2010}, a \emph{task-negative} resting state network, which is more strongly active during idling states than during task performance. The participant's structural connectivity was used to estimate a structural model in the simulator, and its functional connectivity was employed to fine-tune phase and model evaluation.

\subsection{Simulation setup}

For the simulation we used brain data of a 66 years old male, chosen by lot from the set of control participants of a previous study \citep{Garces2014}.

{\scriptsize
\begin{center}
\begin{longtable}{|*4{c|}}
\hline
\textbf{\#}&\textbf{Name}\\ 
\hline
1&Left precuneus\\
\hline
2&Right precuneus\\
\hline
3&Left isthmuscingulate\\
\hline
4&Right isthmuscingulate\\
\hline
5&Left inferiorparietal\\
\hline
6&Right inferiorparietal\\
\hline
7&Left superiorfrontal\\
\hline
8&Right superiorfrontal\\
\hline
9&Left middletemporal\\
\hline
10&Right middletemporal\\
\hline
11&Left anteriorcingulate\\
\hline
12&Right anteriorcingulate\\
\hline
13&Left hippocampus\\
\hline
14&Right hippocampus\\
\hline
 \caption{Description of the 7 NOI per hemisphere considered for the connectomes, obtained from the \emph{Freesurfer} \citep{Fischl2012} cortical parcellation in 66 regions \citep{Desikan2006} such as in \citep{Hagmann2008}.}
 \label{TAB:NodDef}
\end{longtable}
\end{center}
}

To model spontaneous activity, we adjusted the LIFL model to emulate the behaviour of real pyramidal neurons, with a maximum latency of $25 ms$, chosen on the basis of the data reported in the electrophysiology database \emph{Neuroelectro} \citep{Neuroelectro2014}. Neuron's parameters have been set to the following values: $a = 1$, $b = 0$, $c = 0.04$, $t_{arp} = 2$, $D = 0.07$. 
An external excitatory background input is predisposed consisting of spike trains representing the noisy fluctuations observed in vivo, with amplitude chosen in such a way that an isolated neuron displays predominant spiking activity in the alpha band (i.e., 7.5-12 Hz, the frequency range that characterizes real resting state data \citep{Abuhassan2014}). Each node is modeled with $n = 100$ neurons for a total of 1400 neurons. $R=0.8$ as revealed by experimental observations \citep{Izhikevich2004b} and $p = 0.5$ in order to obtain small-world properties. Remaining intra-node connectivity parameters are chosen in such a way that the post-synaptic activity of the nodes preserves its peak in the alpha band, ensuring in the meantime that there is no strict periodicity of individual oscillations, and under the condition $n>>k>>ln(n)>>1$ (as discussed in \ref{SECT:SWWSproc}), obtaining $k=30$ and $\mu_{w_{e,i}} =0.04$. 
Each edge is initialized with a number of connections $N_e / N_{e,max}$ between the considered brain regions, where $N_e$ is equal to the number of streamlines connecting two NOIs reconstructed through DTI \citep[see] [for the method]{Garces2014} and the denominator operates the normalization of the values in the range [0, 1]. Distances between regions $\mu_{\lambda}$ have been obtained considering the Hagmann DSI data set \citep{Hagmann2008}, available as a package of \emph{The Virtual Brain} \citep{SanzLeon2013}.

Final structural model counts 45000 connections, for which plasticity is not active since we are reproducing a static scenario. Simulations have been repeated with two free parameters with the goal to match the resulting FC profile with the real MEG data: conduction velocity $v$, which has been varied in the neighborhood of the best-matching value reported in literature \citep[i.e., 5.1 m/s, see][]{Cabral2014, Nakagawa2014}, and interconnection weights $\mu_{\omega}$ , which we varied in a range that ensures interaction between the nodes but without altering significatively the power spectrum previously set (i.e.,  [0.045, 0.09]). To extract a source-space MEG comparable signal from the model, the \emph{burning event vector} is subsequently imported in Matlab where the events are collected in contiguous bins of 1 ms of simulated time, and the simulated time-series are calculated by summing up all synaptic pulses, as in \cite{Nakagawa2014}. 

From the simulated activity, for each combination of $v$ and $\omega$ we discarded initial transients and extracted 5 segments of 6 \emph{s} of activity. Then we processed the trials of simulated activity in the same way of real MEG source space signal, with the method described in \citep{Garces2014}. 

Finally, for both simulated and real signal we calculated the \emph{amplitude envelope correlation} FC index \citep{Brookes2011} between all pairs of nodes considered. 

The comparison between MEG and model FC matrices is calculated through the Pearson's correlation coefficient $r$ between the strictly upper triangular parts of alpha band FC values across all links connecting the 14 regions of interest (as in \cite{Cabral2014} and \cite{Nakagawa2014}), after the application of the Fisher-Z transform to the FC measures, due to the non-additivity of correlation coefficients \citep{Zimmerman2003}.

\begin{figure}
\centering
\includegraphics[width=0.98\textwidth]{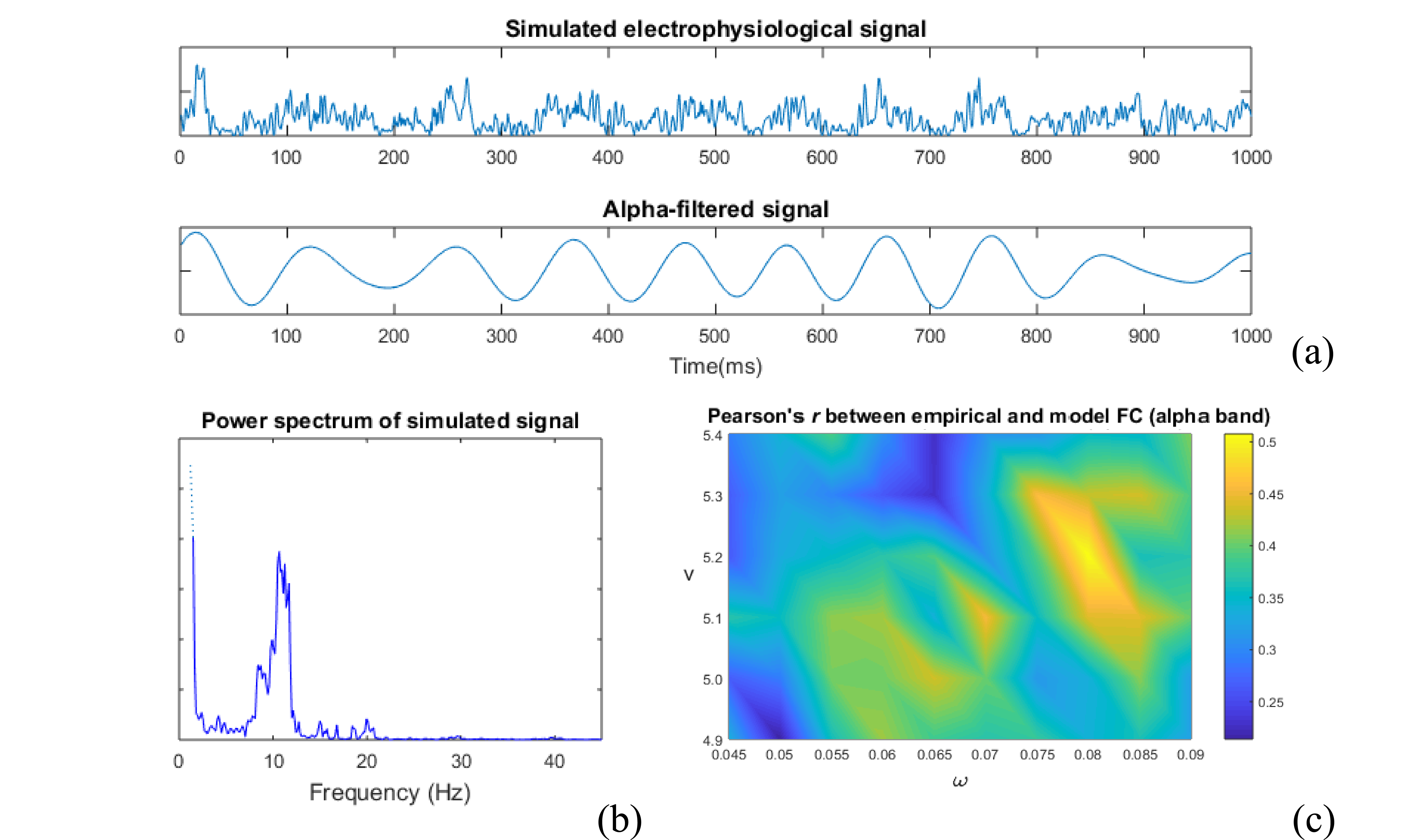}
 \caption{One second of simulated electrophysiological post-synaptic signal extracted from one node of the model with related alpha-filtered counterpart (a); power spectrum obtained from one of the trials (b); (c) represents the Pearson correlation coefficient between the simulated and the empirical FC as a function of simulation parameters v and $\omega$. The figures have been obtained throught the Matlab{\textregistered} script available on the FNS website.} 
\label{FIG:Results}
\end{figure}

The model shows the best agreement with experimental data for the \emph{optimal values} $\omega = 0.080, v=5.2$, reaching an average correlation of $r = 0.51$ between the empirical and simulated FC profiles in the alpha band (Fig. \ref{FIG:Results}) considering the overall set of trials composing the time series, a result that is more than satisfactory if compared to similar studies.
Interestingly, although the groups have been tuned such that isolated neurons display predominant spiking activity in the alpha band, the final model (where the groups are interconnected through weights and delays) shows notable correlation values between MEG and model FC values also in other bands. In facts, using the optimal values  $\omega=0.085$ and $v=5.2$, we have that $r > 0.35$ both in theta and beta bands, indicating the presence of multi-rhythmic activity, in agreement with real data (the complete set of simulation results is published on the Github page of FNS).

For the synthesis of MEG-like signals from the model, two major simplifications have been made:

- regarding the spatial organization of cortical neurons, we consider that pyramidal cells contribute the most to signal generation, taking into account that it is sufficient to reproduce the post-synaptic currents correctly, as shown in previous work \citep [see][for an in-depth discussion on these aspects] {Mazzoni2008};

- the resulting signal more directly corresponds to a simulated LFP then to MEG. Nevertheless, a good correspondence between them has been reported \citep{Nakagawa2014, Zhu2009} due to the fact that both signals arise from the same process (i.e., post-synaptic currents) \citep[]{Buzsaki2012, Susi2018Ch}.

Although the comparison shows that the model is successful in reproducing the FC strength measured in real resting state MEG data, FNS offers many possibilities for improving the model in order to achieve a better comparison, i.e., to consider a larger brain network, to use of a higher parcellation resolution, or to enhance diversity among the modules on the basis of real data.

\subsubsection{Performance}

Considering the example described above, the average time needed for generating a segment of $6s$ of simulated activity is about $100s$ of real time on an \emph{Intel(R) Core(TM) i7-7700CPU @ 3.60GHz }(ram 16GB). The time needed for the initialization phase is about $2s$; the remaining time is for the simulation phase.
While the duration of the first phase depends on the network size only (number of neurons and connections of the network), the duration of the second phase also depends on the value of minimum inter-node length of the network. This time can be reduced by increasing the \emph{maximum heap size} in Java (set to $8GB$ in these simulations).
As explained in Sect. \ref{SECT:OtpS}, during the simulation FNS transcribes $\mathbf{FV}$ and $\mathbf{BV}$ of the selected NOI on the two files \emph{firing.CSV} and \emph{burning.CSV}; the time required to do these operations depends on the number of the selected NOI (14 in this case).

\section{Discussion}

Dynamic models of brain networks can help us to understand the fundamental mechanisms that underpin neural processes, and to relate these processes to neural data. Among the different existing approaches, SNN-based brain simulators allow the user to perform a structure-function mapping at the level of single neurons/synapses, offering a multi-level perspective of brain dynamics.
Here we have presented \emph{FNS}, the first neural simulation framework based on the LIFL model, which combines spiking/synaptic neural modelling with the event-driven simulation technique, able to support real neuroanatomical schemes. FNS allows us to generate models with heterogeneous regions and fibre tracts (initializable on the basis of real structural data), and synaptic plasticity; in addition, it enables the introduction of various types of stimuli and the extraction of outputs at different network stages, depending on the kind of activity to be reproduced. With the aim of showing to the reader a simulation example, we have synthesized a subject-specific brain model sized on real structural data, and analyzed the network spontaneous activity; the comparison with real MEG data shows that FNS is able to reproduce satisfactorily the patterns of neuromagnetic brain sources. In addition, although the nodes have been tuned to oscillate prominently in the alpha band, in the final model there are quite relevant correlation values between MEG and model FC also in other bands (as theta and beta bands), testifying the presence of multirhythmic activity, in agreement with experimental data. This indicates that the spike latency feature could be a key-aspect to reproduce cross-frequency interactions.\\
FNS is downloadable for free and allows the researcher to face realistic simulations, with limited time and budget. The current version of the software is written in Java$ ^{\textregistered}$ and the parallelization is currently implemented as a \emph{multi-threaded} Java standalone application, taking advantage that Java presents specific features oriented to the optimization of memory resources (i.e., the \emph{Java Garbage Collector}), and native support for parallel computation \citep{JavaForkJoin}. 
Future improvements include:
\begin{itemize}
\item translation of our framework to hybrid CPU-GPU technologies (e.g., \emph{CUDA}) or cloud computing scenarios (e.g., \emph{MapReduce});
\item development of an user-friendly interface and improvement of the compatibility with existent FC estimation tools (e.g., \emph{Hermes} \citep{Niso2013});
\item development of a version of FNS based on low level programming languages (e.g., \emph{C} or \emph{GO}) to improve performances and compatibility with \emph{high performing computing} platforms;
\item considering the use of look-up tables \citep{Reutimann2003, Naveros2017} to characterize neuronal dynamics of the LIFL model, for further speeding up simulations. In this process attention must be paid to the insertion of timing errors \citep{Ros2006} to not to compromise some \emph{temporal coding}-based LIFL applications, as \cite{Susi2018}.
\end{itemize}
Although a simulation test has been conducted and commented, this document is not intended as user guide, but as an explanation of the mathematical operation and structure of the simulator. The reader can find the software package and technical documentation on the FNS website: \url{www.fnsneuralsimulator.org} \\


\section{Supplementary material}

\subsection*{Appendix A: LIFL Features}

\renewcommand{\theequation}{B\arabic{equation}}
\setcounter{equation}{0}

\renewcommand{\thefigure}{B\arabic{figure}}
\setcounter{figure}{0}

\begin{figure}
\centering
\subfloat[][]{\includegraphics[width=0.7\textwidth]{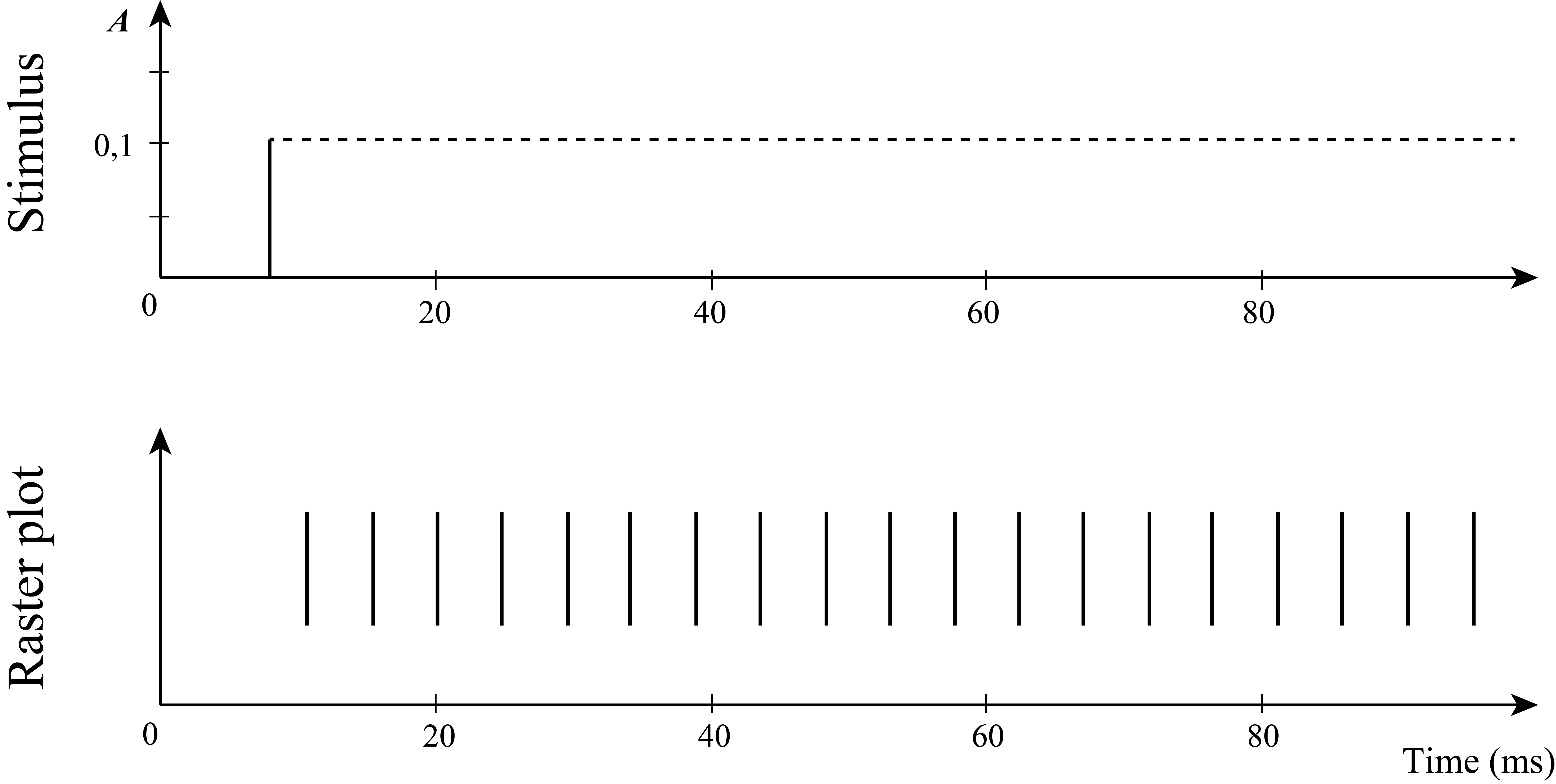}\label{FIG:DC}}
\qquad
\subfloat[][]{\includegraphics[width=0.7\textwidth]{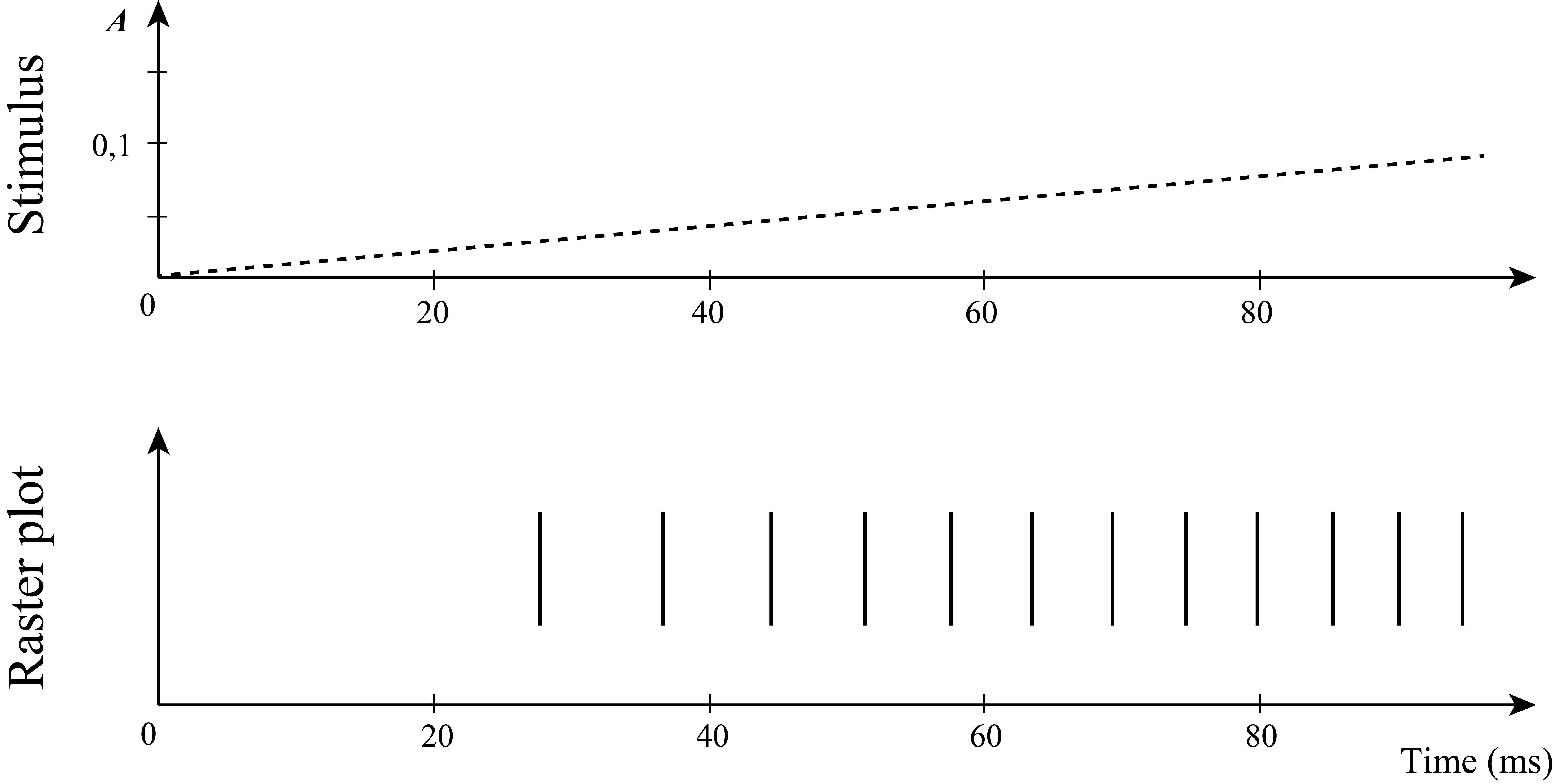}\label{FIG:Ramp}}
\caption{(a): \emph{Tonic Spiking} behavior of LIFL neuron (basic configuration). 
Top: \emph{DC} input stimulus with $A_c = 0.1$. 
Bottom: raster plot (firing activity) of the stimulated neuron, with respect to (simulated) biological time.\\
(b): \emph{Class 1 Excitability} behavior of LIFL neuron (basic configuration). 
Top: \emph{ramp} input stimulus with slope $= 0.001$.
Bottom: raster plot (firing activity) of the stimulated neuron, with respect to (simulated) biological time.\\
Note that, in order to introduce the external inputs to the event-driven system, we used discrete versions of DC and ramp signals, sampled at constant intervals of $dt$ = 0.1 model time}
\label{FIG:StimulatedLIFL}
\end{figure}

LIFL neuron supports natively the following neurophysiological properties: \emph{integrator}, \emph{spike latency}, \emph{tonic spiking} and \emph{class 1 excitability}, obtained by the implementation of the neuron equations in MATLAB environment. Among these:

\begin{itemize}
\item {\emph{Tonic Spiking} takes place when a neuron fires a continuous spike train when stimulated through a DC current input \citep{Izhikevich2004}. We show this property stimulating a single neuron with a constant spike train (i.e., a discretized DC current input) with amplitude $A_c$. The raster plot of the spiking neuron activity (i.e., the output neuron response) is reported in Fig. \ref{FIG:StimulatedLIFL}. Note that the firing frequency of the neuron is constant.}

\item {\emph{Class 1 Excitability}, exhibited by some cortical neurons, allows the neuron to spike with a frequency that depends on the input strength (ranging from 2 Hz to 200 Hz or more), including the fire at low-frequency when the input is weak. This property allows neurons to encode the input strength into their firing rate \citep{Izhikevich2004}. In Fig. \ref{FIG:StimulatedLIFL}, we show this behavior stimulating a neuron with a ramp input. Of course, in this case the firing frequency of the neuron is not constant.}

\end{itemize}

In order to further improve the realism of the LIFL neuron, some adjustments can be made at the programming level (as done for the \emph{refractory period} and the \emph{tonic bursting}), obtaining other computational features as, for example \emph{mixed mode}\citep{Izhikevich2004}.\\

With regards to the underthreshold decay, FNS gives the possibility to choose among:
\begin{itemize}
\item Linear decay:\\
Assuming $T_l = D\Delta t$, in which $D$ is the \emph{decay parameter}, and $\Delta t$ represents the temporal distance between a couple of consecutive incoming spikes (of course, $D \geq$ 0; for $D = 0$ no decay is applied in passive mode, and the neuron behaves as a \emph{perfect integrator}).\\

\item Exponential decay:\\
Assuming $T_l=S_{p\;_{j}}\cdot(1-e^{- \Delta t/D})$, obtaining for the overall update equation:
\begin {equation}
S_{j} =  A_{_{i}}\cdot W_{_{i,j}} +  S_{p\;_{j}} \cdot e^{\frac{-\Delta t} {D}}
\end {equation}
in which $D$ here represents the classic \emph{time constant}.

\end {itemize}

\subsection*{Appendix B: $T_{r}$ calculation}

\renewcommand{\theequation}{A\arabic{equation}}
\setcounter{equation}{0}

\renewcommand{\thefigure}{A\arabic{figure}}
\setcounter{figure}{0}

\begin{figure}[h]

\centering
\includegraphics[width=0.8\textwidth]{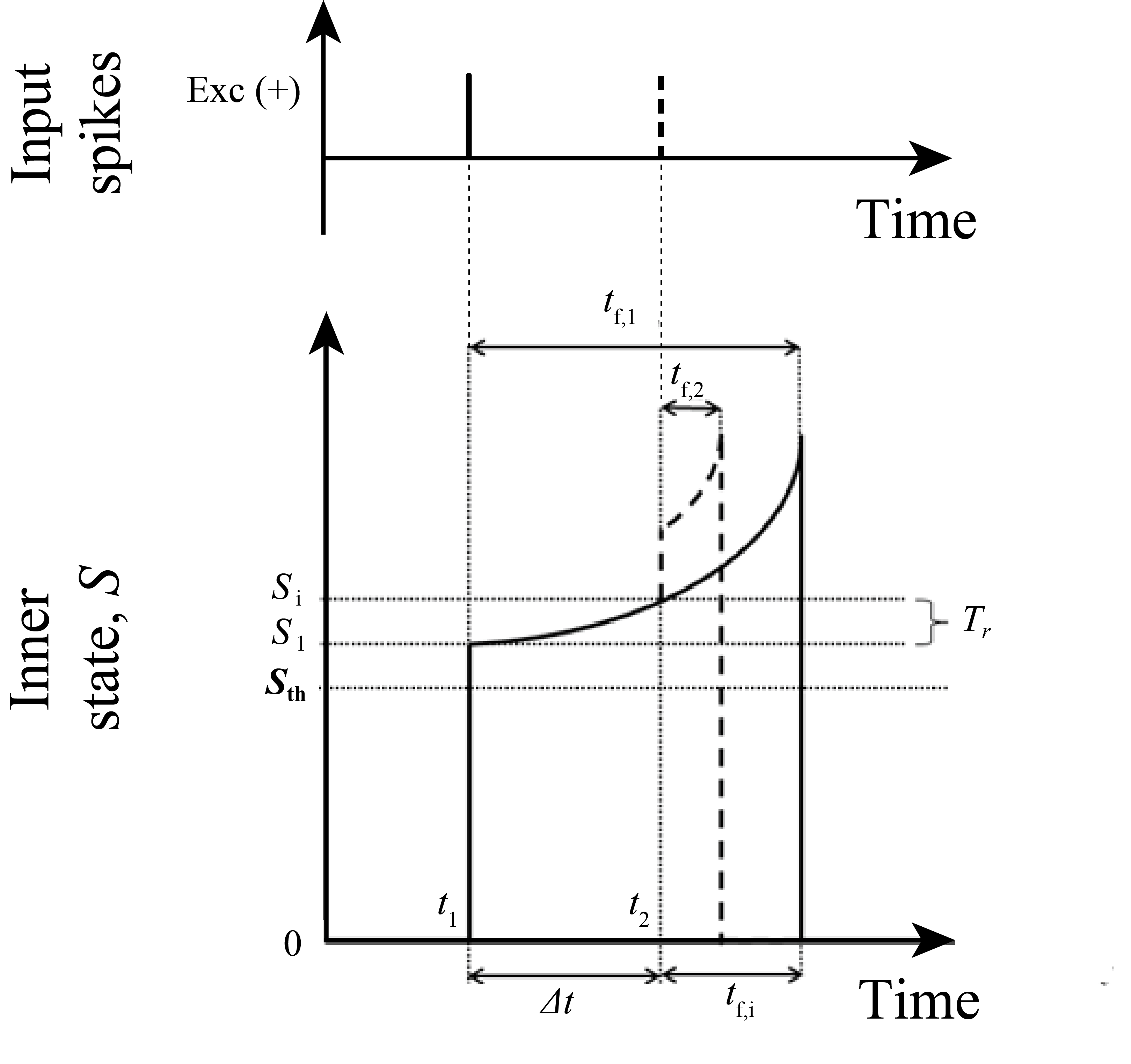}
 \caption{Representation of $T_r$. LIFL neuron in active mode is characterized by a spontaneous growth of $S$. If a pulse arrives before the actual spike generation, $S$ is modified and the $t_{f}$ will be recalculated. The recalculation considers the intermediate state $S_i$, i.e., the neuron state at the time the pulse arrives.} 
\label{Fig:S1_TrCalculation}
\end{figure}

Referring to Fig. \ref{Fig:S1_TrCalculation}, at the time the neurons inner state is altered from a second input (here excitatory, but non influencial to calculation purposes), the \emph{intermediate state} $S_i$ is determined, and then $T_{r}$ is calculated. 

In event-driven, network elements are updated only when they receive or emit a spike. Once an input spike arrives in active mode, the $S_i$ is calculated on the basis of the time remaining to the spike generation.

Referring to the generic inner state $S_{i}$ the firing equation is:

\begin{equation}\label{EQ:tfi}
t_{f,i}=\frac{a}{S_i-1}-b
\end{equation}

We define:

\begin{equation}
\Delta t=t_2-t_1
\end{equation}
where $t_1$ and $t_2$ represent the arrival instants of the synaptic pulses to the considered neuron. Then:

\begin{equation}\label{EQ:tfiNEW}
t_{f,i}=t_{f,1}-\Delta t 
\end{equation}

Rearranging Eq. \ref{EQ:tfi}, we obtain:
\begin{equation}\label{EQ:SiNEW}
S_i=\frac{a}{t_{f,i}+b} +1
\end{equation}

Now we combine Eq. \ref{EQ:tfiNEW} with Eq. \ref{EQ:SiNEW}
\begin{equation}\label{EQ:SiNEW2}
S_i=\frac{a}{t_{f,1}-\Delta t + b}+1
\end{equation}

By defining
\begin{equation}\label{EQ:Treq}
T_r = S_i - S_1
\end{equation}

where
\begin{equation}\label{EQ:S1}
S_1=\frac{a}{(t_{f,1} + b)} + 1
\end{equation}

and putting Eq .\ref{EQ:SiNEW2} and \ref{EQ:S1} in \ref{EQ:Treq}, we obtain:

\begin{equation}
T_r=\frac{a}{t_{f,1}-\Delta t + b}-\frac {a}{t_{f,1} + b}
\end{equation}

that can be rearranged as

\begin{equation}\label{EQ:DeltaSNEW}
T_r=\frac{a \Delta t}{(t_{f,1} + b -\Delta t) (t_{f,1}+b) }
\end{equation}

Note that we are interested in determining an intermediate state; this implies that we consider the second synaptic pulse only if its timing (i.e., $t_2$) falls before the spike occurs. This gives us:
\begin{equation}\label{EQ:condition}
\Delta t < t_{f,1}
\end{equation}
thus we do not have restrictions from the denominator of \ref{EQ:DeltaSNEW}.

The relation \ref{EQ:DeltaSNEW} can be generalized to the case as more input modify the firing time; then, we can write
\begin{equation}
T_r =S_{ic}-S_{ip}=\frac{a \Delta t_i}{(t_{f,ip} + b -\Delta t_i)(t_{f,ip}+b) }
\end{equation}

with

\begin{equation}
\Delta t_i=t_{ic}-t_{ip}
\end{equation}

where the subscript $ip$ stays for \emph{intermediate-previous} and $ic$ for \emph{intermediate-current}. 

We can also make explicit the dependence of $T_r$ from the previous state, by inverting $t_{f,ip}$ trough Eq. \ref{EQ:tfi}, obtaining:

\begin{equation}\label{EQ:DeltaSNEW_doubleintermediate}
T_r=\frac{(S_{ip}-1)^2 \Delta t}{a-(S_{ip}-1)\Delta t}
\end{equation}

Obviously, the same considerations on the arrival time of the second pulse remain valid, thus we do not have restrictions imposed by the denominator of \ref{EQ:DeltaSNEW_doubleintermediate}.

\subsection*{Appendix C: types of distributions used in the model}

\renewcommand{\theequation}{C\arabic{equation}}
\setcounter{equation}{0}

\renewcommand{\thefigure}{C\arabic{figure}}
\setcounter{figure}{0}

In FNS Gaussian distributions are implemented both for the initialization of intra-module weights of modules (one set for excitatory and one set for inhibitory) and inter-module weights of each edge:

\begin{equation}
  f(W) =\frac{1}{{\sqrt{2 \pi {\sigma_{W}} ^2 } } }  exp{\frac {(W-\mu_{W})^2} {-2 {\sigma_{W}} ^2}} \; \label{EQ:Gw}
\end{equation}

where $\mu_{W}$ is the mean, and $\sigma_{W} ^{2}$ is the variance of the distribution. In the formula, W is intended to represent $A$ (distribution of intra-module weights) or $\omega$ (distribution of intermodule weights).

In time, a gamma distribution is implemented for inter-module lengths, which reflects on a gamma distribution of delays. Such gamma distribution is characterized from parameters $\mu_{\lambda}$ (i.e., \emph{mean parameter}) and $\alpha_{\lambda}$ (i.e., \emph{shape parameter}). If we call $\lambda$ the (axonal) delay, the probability density function of a gamma distribution can be written as:

 \begin{equation}
    f(\lambda) =  \lambda^{\alpha_{\lambda}-1} \frac {exp(-\lambda \alpha_{\lambda}/\mu_{\lambda})}{(\mu_{\lambda}/\alpha_{\lambda})^{\alpha_{\lambda}} \Gamma (\alpha_{\lambda})} \;   \label{GammaFunct1}
 \end{equation}
 
Note that $\mu_{\lambda}$ can be defined as: 

 \begin{equation}
   \mu_{\lambda} =\alpha_{\lambda} \theta\; \label{GammaFunct2}
 \end{equation}
 
where $\theta$ is known as the \textit{scale parameter}.\\

Note that with the parameter $\alpha_{\lambda}$ is possible to control the type of distribution (low $\alpha_{\lambda}$ values lead toward the exponential distribution; high $\alpha_{\lambda}$ values lead toward the Dirac distribution); the more $\alpha_{\lambda}$ is high, the more the distribution \emph{mode} approaches $\mu_{\lambda}$ (from the left). 

\subsection*{Appendix D: Pseudo-code procedure}

\renewcommand{\theequation}{D\arabic{equation}}
\setcounter{equation}{0}

\renewcommand{\thefigure}{D\arabic{figure}}
\setcounter{figure}{0}

\begin{description}
\item [$\cdot$]let \verb|spike_queue| be the list of all spikes generated by any pre-synaptic neuron sorted in ascending order of spike time (the first spike in the list is the one with the least spike time);
\item [$\cdot$]let \verb|outer_burning_event|: the list of spikes to be delivered to post-synaptic neurons belonging to outer nodes. At each synchronization:
\\  1. each entry of this list is read and sent to the thread executing the routine for the right node;
\\  2. after each item in the list has been sent, the whole list is cleared.
\item [$\cdot$]let \verb|current_time| be the current simulated time;
\item [$\cdot$]let \verb|split_stop_time| be the stop time for the current BOP simulation;
\item [$\cdot$]let \verb|final_stop_time| be the simulated time at which the whole simulation must stop;
\item [$\cdot$]let \verb|run_burning_routine(s*)|  be the routine which calculates all the burning events caused by the fire event hold by the spike \verb|s*|: during this procedure, all the burning events involving post-synaptic neurons of outer nodes are stored to a special \verb|outer_burning_event|;
\item [$\cdot$]let \verb|send_fires_for_outer_nodes()| be the routine which sends the spikes stored in \verb|outer_burning_event| to post-synaptic neurons in outer nodes;
\item [$\cdot$]let \verb|update_incoming_spikes_queue()| be the routine which updates the \verb|spike_queue| with the spikes coming from outer nodes and targeting post-synaptic neurons of the present node before beginning the next BOP simulation.

\end{description}

\begin{verbatim}
while true:
   while current_time < split_stop_time:
      s* = spike_queue.pop()
      run_burning_routine(s*)
      if current_time >= split_stop_time:                      (1)
         if split_stop_time >= final_stop_time:                (2)
            return
         send_fires_for_outer_nodes() 
      wait_until(current_time < split_stop_time)               (3)
  update_spikes_queue()


(1) end of the BOP: send the fires to the outer nodes
(2) end of the last BOP: end of the node simulation
(3) stops the thread until split_stop_time is updated 
    with the next BOP stop time

\end{verbatim}

\subsection*{Information Sharing Statement}
Software, simulation results and new network models are available at the \\
FNS official website [www.fnsbrainsimulator.org] and GitHub \\
page [https://github.com/fnsneuralsimulator].\\

\section*{References}

\bibliographystyle{elsarticle-harv}

\end{document}